\documentclass[prb,aps,showpacs,twocolumn,preprintnumbers,
amsmath,amssymb,superscriptaddress,longbibliography]{revtex4-2}
\usepackage[english]{babel}
\usepackage{amsmath,amssymb,amsfonts}
\usepackage{graphicx}
\usepackage[colorlinks=True,linkcolor=red,citecolor=blue,urlcolor=blue]{hyperref}

\usepackage{booktabs}
\usepackage[dvipsnames]{xcolor}
\usepackage{braket}
\usepackage{bm}
\usepackage{bbm}

\usepackage{enumitem}

\usepackage{braket}
\usepackage{float}
\usepackage{multirow}
\usepackage{longtable}
\usepackage[normalem]{ulem}
\usepackage{array}
\usepackage{makecell}
\usepackage{subfigure}
\usepackage{booktabs}
\usepackage{multirow}
\usepackage{graphicx}% Include figure files
\graphicspath{{NewFigures/}}
\usepackage{dcolumn}% Align table columns on decimal point
\usepackage{bm}% bold math
\usepackage{longtable}
\usepackage{xcolor}
\newcolumntype{C}{>{$}c<{$}}

\usepackage{listings}
\usepackage{color}

\definecolor{dkgreen}{rgb}{0,0.6,0}
\definecolor{gray}{rgb}{0.5,0.5,0.5}
\definecolor{mauve}{rgb}{0.58,0,0.82}

\def\br{\mathbf{r}}
\def\bk{\mathbf{k}}
\def\id{\mathbbm{1}}

\newcommand{\abs}[1]{\left\lvert #1 \right\rvert}

\renewcommand{\Re}{\mathop{\mathrm{Re}}}
\renewcommand{\Im}{\mathop{\mathrm{Im}}}

\allowdisplaybreaks
\DeclareMathAlphabet{\zc}{OT1}{pzc}{m}{it}

\def\ket#1{|#1\rangle }

\begin{document}

\title{Beyond symmetry protection: Robust feedback-enforced edge states in non-Hermitian stacked quantum spin Hall systems}

\author{Mengjie Yang}
\email{mengjie.yang@u.nus.edu}
\affiliation{Department of Physics, National University of Singapore, Singapore 117551, Singapore}

\author{Ching Hua Lee}
\email{phylch@nus.edu.sg}
\affiliation{Department of Physics, National University of Singapore, Singapore 117551, Singapore}

\date{\today}

\begin{abstract}
  Conventional wisdom holds that, in the simplest time-reversal-symmetric setting, strongly coupling two QSH layers yields a trivial $\mathbb Z_2$ phase and no protected topological edge states. We demonstrate that, in a regime with intermediate inter-layer coupling (neither in the strong or weak coupling regimes) and competitive non-Hermitian directed amplification, bulk modes are rendered with negligible gain while arbitrary bulk excitations inevitably accumulate into robust helical edge transport modes --- without relying on any symmetry protection. Our feedback-enforced mechanism persists over broad parameter ranges and remains robust even on fractal or irregular boundaries. These findings challenge the traditional view of stacked QSH insulators as inevitably trivial, and open up new avenues for designing helical topological devices that exploit feedback-enforced non-Hermitian engineering, instead of symmetry-enforced robustness.
\end{abstract}

\maketitle

\section{Introduction}

Topological insulators~\cite{hasan2010colloquium,qi2011topological,bernevig2013topological,asboth2016short,bergholtz2021exceptional}, particularly quantum spin Hall (QSH) systems~\cite{kane2005quantum,bernevig2006quantum,knez2011evidence,kane2005z,konig2007quantum,roth2009nonlocal}, host helical edge states protected a $\mathbb{Z}_2$ topological invariant~\cite{kane2005z,konig2007quantum,roth2009nonlocal,moore2007topological,gu2016holographic,roy2009z,qin2022light,OCHKAN2026100630}. Conventionally, when two QSH layers are brought into strong proximity, inter-layer tunnelling opens a hybridization gap that annihilates the counter-propagating helical modes, rendering the composite stack $\mathbb Z_{2}$-trivial~\cite{zhang2009topological,liu2010model,liu2010oscillatory,yoshimi2015quantum,hui2015bulk,shoman2015topological,song2015probing,hsieh2016bulk,panas2020bulk,stuhler2022effective} (See, e.g. Refs.~\cite{michetti2012tunable,michetti2013devices,baum2015coexisting,choi2023stacking} for scenarios where appropriate stacking symmetries or biases preserve non-trivial topology.).

Yet, stacking QSH insulators in an appropriate intermediate coupling regime can lead to a wealth of new phenomena~\cite{zhang2009topological,liu2010model,shan2010effective,knez2011evidence,qian2014quantum,zhong2023towards,kang2024double,kang2024evidence}. Here, the coupling is neither too strong--where hybridization destroys the helical edge states--nor too weak--where the layers remain effectively decoupled. This includes large-gap QSH insulators in 2D transition metal dichalcogenides~\cite{qian2014quantum,tang2017quantum,wu2018observation,shi2019imaging}, Chern number hierarchy in topological flat bands~\cite{liu2019quantum,lee2014lattice,wang2022hierarchy}, the potential for layer-selective QSH channels in self-assembling vdW materials~\cite{zhong2023towards}, and the emergence of double QSH phases in twisted Moiré materials, namely, the coexistence of two pairs of helical edge states~\cite{kang2024double,kang2024evidence}.

Central to our construction is the non-Hermitian skin effect (NHSE)~\cite{lee2016anomalous,alvarez2018non,yao2018edge,kunst2018biorthogonal,lee2019anatomy,yokomizo2019non,okuma2020topological,lin2023topological,okuma2023non,longhi2019topological,kawabata2019symmetry,lee2019hybrid,song2019non,borgnia2020non,helbig2020generalized,li2020critical,longhi2020non,lee2020unraveling,zou2021observation,zhang2021observation,yang2022concentrated,zhang2021tidal,li2022non,xue2021simple,xue2022non,longhi2022self,gu2022transient,shen2022non,jiang2023dimensional,xiong2024non,shimomura2024general,yang2024percolation,gliozzi2024many,yang2025non,li2025phase,yoshida2024non,wang2026exactsolutionsnonreciprocalsuschriefferheeger}, whereby non-reciprocity causes an extensive number of bulk eigenstates to accumulate near a boundary under open boundary conditions, producing a nonperturbative sensitivity to boundary conditions that invalidates the conventional Bloch description and qualitatively reconstructs the bulk spectrum~\cite{yao2018edge,kunst2018biorthogonal,lee2019anatomy,yokomizo2019non,okuma2020topological,helbig2020generalized,xue2026non}. In two dimensions, the skin accumulation can be directional or occur at boundaries of higher codimension~\cite{lee2019hybrid,helbig2020generalized,song2020two,zou2021observation,zhang2021observation,zhang2021tidal,jiang2023dimensional,manna2023inner,xiong2024non,yang2026reversing}. Importantly, NHSE-induced boundary accumulation is distinct from the pre-existing topological edge states of a Chern system: states originating from the bulk bands may acquire strong boundary localization while remaining spectrally distinct from the in-gap topological edge branches.

In this work, we demonstrate that non-Hermiticity acting in this intermediate coupling regime can lead to a new form of robustness in the edge transport, departing from the conventional paradigm of symmetry-protected topology. Our mechanism hinges on strategically interfering the directed NHSE amplification in different Chern layers, such that bulk diffusion acquires essentially no gain and any bulk excitations are spontaneously funneled into robust, exponentially amplified chiral edge transport. 
Saliently, it operates with intermediate-strength couplings that would instantly destroy critical NHSE~\cite{li2020critical,liu2020helical,arouca2020unconventional,liu2020helical,guo2021exact,yokomizo2021scaling,rafi2022system,rafi2022critical,qin2023universal,yang2024percolation,rafi2025critical,yang2026nonlocalityinducedcriticallengthhierarchynonhermitian} scalings.
In stark contrast to previous approaches whose edge-mode robustness fundamentally depends on specific symmetries such as $\mathcal{PT}$ symmetry~\cite{yuce2018pt,song2020pt,sone2020exceptional,slootman2024breaking,fritzsche2024parity}, our proposed mechanism is inherently symmetry-agnostic, valid even if no symmetry exists between the constituent topological layers~ (In the Hermitian, time-reversal-invariant limit ($\gamma=0$), an $N$-layer QSH stack is classified by a $\mathbb{Z}_2$ index $\nu_{\rm tot}=N\bmod 2$, so that a four-layer stack is $\mathbb{Z}_2$-trivial. The edge-dominant amplification we study is, however, an intrinsically non-Hermitian effect that vanishes as $\gamma\to 0$. In the regime $\gamma\neq 0$ (where conventional TRS is broken), our construction is more naturally viewed as a strongly coupled stack of four Chern layers with nonzero Chern numbers; we do \emph{not} require any $\mathbb{Z}_2$ topology, but instead rely on non-Hermitian feedback competition between these layers to obtain an almost-real bulk spectrum together with strongly amplified edge channels.). Consequently, our edge states persist even under strong symmetry-breaking perturbations and remain resilient along highly irregular boundaries, thereby uncovering a previously unidentified pathway to symmetry-agnostic robustness in topological transport.

\begin{figure*}
  \centering
  \includegraphics[width=0.95\textwidth]{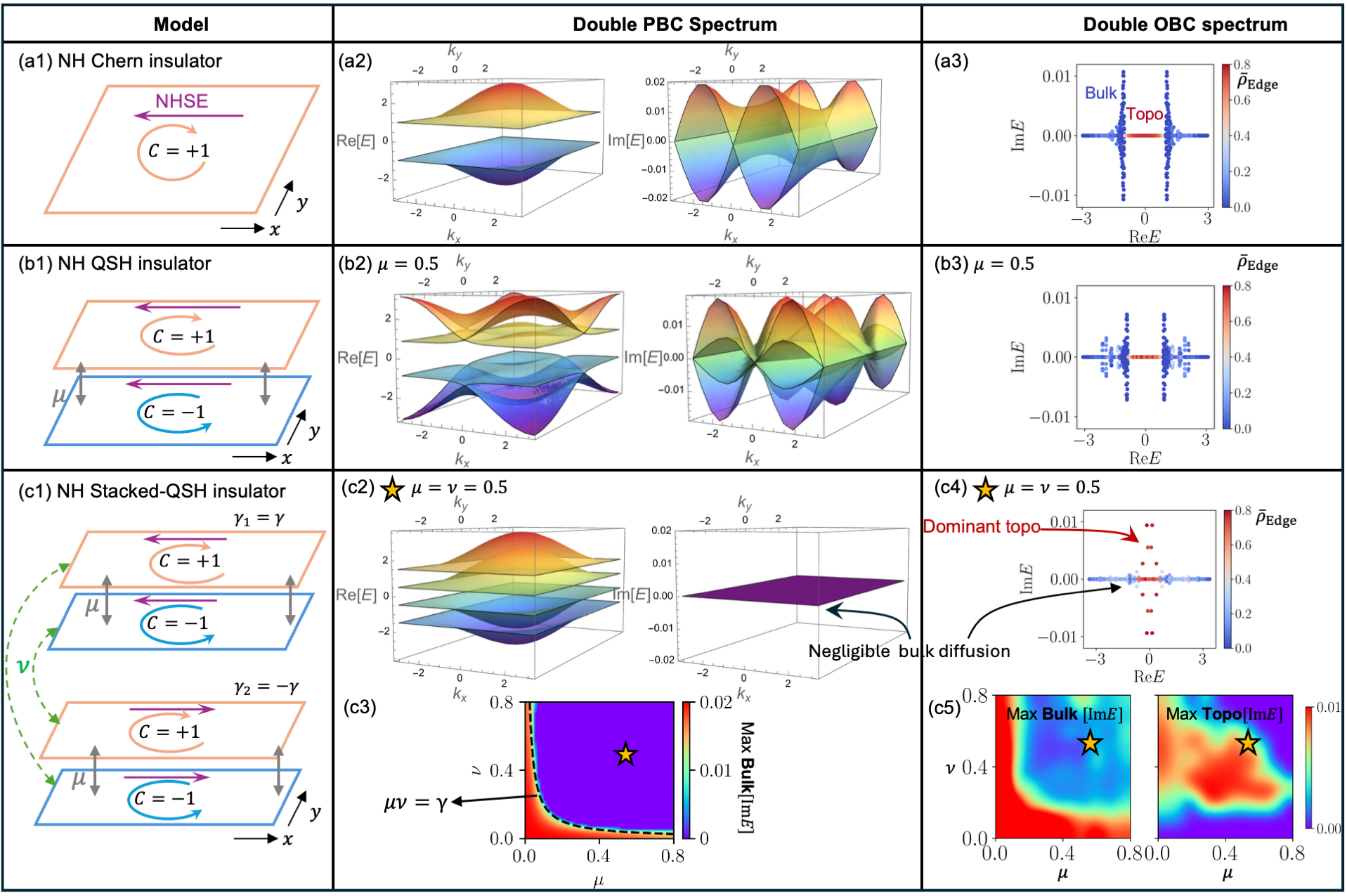}
  \caption{Scheme for achieving greatly enhanced OBC edge Im$E$ and essentially gainless bulk Im$E$ for robust feedback-enforced edge transport.
    (a1) An individual Non-Hermitian Chern insulator $H_\text{Ch}$ layer possesses undesirably essentially gainless edge amplification: Its (a2) PBC bulk and (a3) OBC bulk bands (blue) are both complex, while topological edge states (red) have real energies. OBC eigenstates $\psi_n(\br)$ are colored by edge occupation ratio [$\bar\rho_{\text{Edge}}\equiv\rho_{\text{Edge}}[\psi_n(\br)]$ from Eq.~\eqref{eq:edge_ratio_general}].
    (b1) $\mu$ couples two non-Hermitian Chern layers of opposite chirality but parallel NHSE direction, yielding a non-Hermitian QSH bilayer [Eq.~\eqref{eq:QSH}]: However, its (b2) PBC bulk and (b3) OBC bulk bands (blue) are still complex, while helical edge states (red) are still real.
    (c1) Our 4-layer non-Hermitian stacked-QSH system [Eq.~\eqref{eq:biQSH}] consists of two QSH copies with opposite NHSE coupled by $\nu$. Suitable parallel and anti-parallel NHSE couplings $\mu$ and $\nu$ can achieve real (c2) PBC bulk and (c4) OBC bulk states (blue), accompanied by dominant complex OBC topological edge states that would be strongly amplified (red).
    (c3, c5) PBC bulk and OBC bulk/edge $\max\left[\operatorname{Im}E_{\text{stacked-QSH}}\right]$ in the $(\nu,\mu)$ parameter space, characterized by a large nearly real-bulk regime ($\operatorname{Im}E_{\text{stacked-QSH}}=0$) that partially coincides with an edge enhancement ($\operatorname{Im}E_{\text{stacked-QSH}}>0$) regime (see Supplementary Secs.~S1 and S2).
    As such, edge modes can dominate over bulk modes with intermediate inter-layer coupling $\nu$ -- neither too strong (which would cause cancellation of helical states and yield a trivial phase), nor too weak (failure to establish effective 4-layer connectivity).
  All constituent Chern layers are computed with $m_1=m_2=m=1$ and $\gamma_1=-\gamma_2=\gamma=0.02$, with lattice size $L_x\times L_y=40\times 10$.}
  \label{fig:fig1}
\end{figure*}

\section{Results}
\subsection{Scheme for robust amplificative edge modes}
In non-Hermitian contexts, edge states can be extremely robust not just from topological protection, but also because they experience stronger gain than the bulk. This requires a mechanism that suppresses the imaginary energy Im$E$ for the bulk states relative to the edge states. Below, we show how to achieve this in a highly versatile manner by coupling four Chern (i.e. two QSH) layers.

Our approach involves stacking two QSH insulators that experience opposite NHSE, as can be physically achieved through flux and gain/loss, not necessarily with asymmetric couplings~\cite{li2022gain,wu2022flux,li2023loss,liu2023spin,li2025observation,liuobservation,wu2021floquet}.
Each QSH system, labeled $j=1,2$, is made up of two coupled Chern layers [Fig.~\ref{fig:fig1}(a1)] with opposite chirality but common NHSE direction. A simplest specific model is given by [Fig.~\ref{fig:fig1}(b1)]
\begin{equation}
  H_{\text{QSH}}^{(j)}=
  \begin{pmatrix}
    H_{\text{Ch}}(\bk,m_1, +\gamma_j) & \mu \id  \\
    \mu \id & H_{\text{Ch}}(\bk,-m_2, -\gamma_j)
  \end{pmatrix}, \label{eq:QSH}
\end{equation}
where $H_{\text{Ch}}(\bk, m, \gamma_j) = \left(m + \cos k_x + \cos k_y\right) \sigma_x + \left(i\gamma_j + \sin k_x\right) \sigma_y + \sin k_y \sigma_z$ describes a non-Hermitian Chern insulator~\cite{kawabata2018anomalous}. Here $m$ is the Dirac-mass parameter that controls the Hermitian Chern phase, while $\gamma_j$ is the non-Hermitian strength; the term $i\gamma_j\sigma_y$ produces unequal effective hoppings along $x$ and hence the $x$-NHSE. Notably, we do not require $m_1=m_2$ (as long as they are in the same topological class), since a symmetrical gap or band structure is not a requisite ingredient.
We refer to $\mu$ as the \textit{parallel helical coupling}, since it connects two Chern insulators exhibiting the NHSE in the \emph{same} direction. For fixed $k_y$, the asymmetric $x$-direction hopping amplitudes are proportional to $m+\cos k_y\pm\gamma_j$. The standard non-Bloch similarity transformation therefore gives the generalized-Brillouin-zone radius
$r_x(k_y)=\sqrt{\left|\frac{m+\cos k_y-\gamma_j}{m+\cos k_y+\gamma_j}\right|}$
and the inverse skin depth
$\kappa_x(k_y)=-\ln r_x(k_y)$. The two Chern layers have the same $x$-NHSE direction and the same $|\kappa_x|$ up to the translation $k_y\rightarrow k_y+\pi$; see Refs.~\cite{yao2018edge,yokomizo2019non,okuma2020topological}.

Shown in Fig.~\ref{fig:fig1}(a2, a3) and (b2, b3) are the Chern and QSH insulators spectra computed, respectively, under periodic boundary conditions in both $x$ and $y$ directions (double PBC) and open boundary conditions in both directions (double OBC), with $m_1=m_2=1$ and $\gamma_j=0.02$ for simplicity, such that the two layers exhibit Chern numbers $C=\pm1$.
With these values, the bulk spectra are always complex under both PBCs and OBCs~ (
The PBC spectrum of $H_{\text{Ch}}$ encloses nonzero area due to the non-reciprocal $\pm i\gamma_j$, while under OBCs, the effective $x$-hoppings become $\sqrt{(1+\cos k_y)^2-\gamma_j^2}$ in the generalized Brillouin zone (GBZ)~\cite{kawabata2018anomalous,mandal2022topological,hou2021topological,ma2023quantum,liu2023spin,ochkan2024non}, which is imaginary for some $k_y$.), while the topological edge modes always remain real because they can be mapped to their Hermitian limits by trivial 1D basis rescalings~\cite{yao2018edge,yokomizo2019non,okuma2020topological,kawabata2020real,hou2021topological,yang2024percolation,qi2024extended}~ (For the rectangular geometry with open boundaries in both $x$ and $y$, the topological edge modes remain strictly real: the real-space Hamiltonian is pseudo-Hermitian and unbroken global $\mathcal{PT}$ symmetry pins the edge eigenvalues to the real axis (see Supplementary Sec.~S7).). Here, the OBC states $\Psi(\bold r)$ are colored according to their edge occupation ratio
\begin{equation}
  \rho_{\text{Edge}}[\Psi] = \frac{\sum_{\bold r \in \text{Edge}}|\Psi(\bold r)|^2 - \frac{A_{\text{Edge}}}{A_{\text{Bulk}}} \sum_{\bold r \in \text{Bulk}}|\Psi(\bold r)|^2}{\sum_{\bold r}|\Psi(\bold r)|^2}, \label{eq:edge_ratio_general}
\end{equation}
$A_{\rm Bulk}$ and $A_{\rm Edge}$ denote the numbers of physical lattice sites in the bulk and edge regions, respectively, with $\mathbf r$ running over the two sublattice sites in each unit cell. We define the edge as the one-unit-cell-wide outermost perimeter, including both sublattice sites of every boundary unit cell. Thus, for an $L_x\times L_y$ lattice,
$A_{\rm Edge}=4L_x+4L_y-8$ and $A_{\rm Bulk}=2(L_x-2)(L_y-2)$.
Here, $|\Psi(\mathbf r)|^2$ denotes the local probability density summed over the layer degrees of freedom. A normalized uniform state satisfies $|\Psi(\mathbf r)|^2=1/(2L_xL_y)$ at every physical lattice site and hence gives $\rho_{\rm Edge}=0$. Complete edge localization gives $\rho_{\rm Edge}=1$, whereas $\rho_{\rm Edge}<0$ indicates bulk-dominant localization. 
Evidently, both the Chern and QSH insulators are by themselves incapable of exhibiting robust edge transport, since it is their bulk states that possess Im$E>0$.

Interestingly, by stacking two such QSH systems, we can make the edge and not bulk states exhibit Im$E>0$ instead. We define the stacked QSH system [Fig.~\ref{fig:fig1}(c1)] as
\begin{equation}
  H_{\text{stacked-QSH}} =
  \begin{pmatrix}
    H_{\text{QSH}}^{(1)} & \nu\sigma_x\otimes \id \\
    \nu\sigma_x\otimes \id & H_{\text{QSH}}^{(2)}
  \end{pmatrix}.  \label{eq:biQSH}
\end{equation}
In addition to the previously introduced parallel helical couplings $\mu$ [gray solid arrows in Fig.~\ref{fig:fig1}(c1)], we also have inter-QSH \textit{anti-parallel helical couplings} $\nu\sigma_x$  [green dashed arrows], which couple opposite Chern layers exhibiting NHSE in \emph{anti-parallel} directions (since they belong to different QSH systems). Here their $\gamma_1,\gamma_2$ just need to be of opposite sign, and we henceforth set $\gamma_1=-\gamma_2=\gamma$ without loss of generality~ (For such models with circular GBZs, we can in general write $\gamma_1=\gamma'+\gamma$ and $\gamma_2=\gamma'-\gamma$, where the common offset $\gamma'$ can be gauged out though a basis rescaling, leaving only the relative non-Hermiticity $\gamma$ as physically significant. ).

Shown in Figs.~\ref{fig:fig1}(c2) and (c4) are examples where the bulk bands of $H_{\text{stacked-QSH}}$ are essentially real. Here ``essentially real'' distinguishes the two boundary conditions. Under double PBC, the bulk spectrum is strictly real when $\mu\nu\ge\gamma$ (Supplementary Sec.~S1). Under finite double OBC, the bulk eigenvalues retain small residual imaginary parts from finite-order bulk corrections and boundary/corner hybridization, as bounded in Supplementary Sec.~S2. For the parameters of Fig.~\ref{fig:fig1}, these residual bulk imaginary parts are much smaller than both the single-layer scale $\gamma$ and the imaginary parts of the amplified edge modes. The edge states (red in (c4)) acquire Im$E>0$ and will be amplified with time. Notably, qualitatively similar behavior persist in an extended parameter space region: as plotted in Figs.~\ref{fig:fig1}(c3) and (c5), the PBC bulk becomes real (purple region on the right side of the dashed curve) as long as $\mu\nu\geq\gamma$ (see Supplementary Sec.~S1), i.e. both $\mu,\nu$ cannot be too small. Similar conclusions apply to the OBC bulk [blue in Fig.~\ref{fig:fig1}(c5); see Supplementary Sec.~S2]. Fortuitously, this parameter region strongly overlaps with that where the OBC topological states (taken to be those within the bulk gap) exhibit the largest Im$E$ (red central region in Fig.~c5) where $\mu,\nu$ are both comparable with the scale of $m_1=m_2=1$. In this ``intermediate coupling'' regime, excitations are hence amplified much more at the edge than in the bulk.
Equivalently, in terms of non-Hermitian spectral topology, $\omega_{\rm bulk}$ and $\omega_{\rm edge}$ characterize two different spectral problems rather than two parts of the same finite-system spectrum. The bulk invariant $\omega_{\rm bulk}$ is the point-gap winding of $\det[H_{\text{stacked-QSH}}(k_x,k_y=0)-E_\ast]$ along periodic $k_x$, whereas $\omega_{\rm edge}$ is the winding of $\det[g_s^{-1}(k_x,E_\ast)]$ for the surface Green's function of a semi-infinite system open along $y$. At $E_\ast=0$, we obtain $\omega_{\rm bulk}=0$ and $\omega_{\rm edge}=2$: the periodic bulk spectrum does not wind around the base energy, while the boundary spectral problem has winding two. This nontrivial boundary winding is accompanied by two edge channels whose associated eigenvalues have larger $\operatorname{Im}E$ than the eigenvalues associated with bulk-localized modes (see Supplementary Sec.~S5).
Note that this edge enhancement can exist even when the OBC spectra of the constituent Chern layers are fully real, as elaborated in Supplementary Sec.~S3.

\begin{figure}
  \centering
  \includegraphics[width=0.5\textwidth]{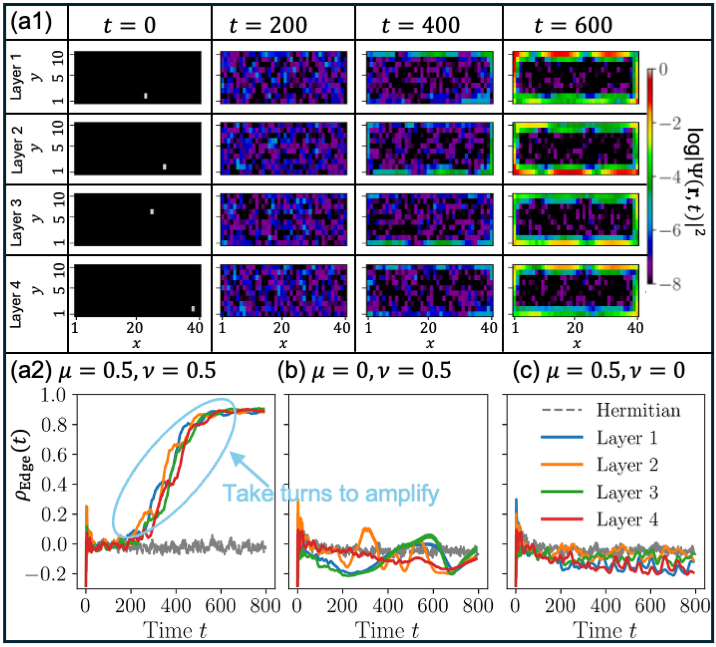}
  \caption{Feedback-enforced chiral edge dynamics in our non-Hermitian stacked-QSH system [Eq.~\eqref{eq:biQSH}] with $\gamma=0.02$, $L_x\times L_y=40\times 10$ and various combinations of couplings $\mu$ and $\nu$, which respectively connect layers with parallel and anti-parallel NHSE [see Fig.~\ref{fig:fig1}(c1)].
  (a1) Propagation of a wavepacket $\Psi(\bold r, t)=\exp(-iH_{\text{stacked-QSH}}t)\Psi(\bold r\notin\text{Edge}, t=0)$ under $\mu=\nu=0.5$ (intermediate coupling regime). The random initial bulk excitations at $t=0$ are gradually redirected into clean chiral edge propagation. (a2) The corresponding feedback-enforced mechanism is evident in the layer-resolved edge occupation ratio $\rho_{\text{Edge}}(t)\equiv \rho_{\text{Edge}}[\Psi(\br, t)]$ [Eq.~\eqref{eq:edge_ratio_general}], where the four curves intertwine and take turns to rise. In contrast, $\rho_{\text{Edge}}(t)\sim 0$ [gray] for the same initial state with non-Hermiticity turned off ($\gamma=0$). Feedback-enforced edge amplification is similarly absent unless both $\mu,\nu$ couplings are present: In (b) $\mu=0.5, \nu=0$ and (c) $\mu=0, \nu=0.5$, the edge occupation ratio $\rho_{\text{Edge}}(t)\approx 0$ in both non-Hermitian and Hermitian cases. }
  \label{fig:fig2}
\end{figure}

\subsection{Feedback-enforced helical edge transport dynamics}
Below, we describe how our robust edge amplification mechanism arises dynamically,
not only rendering non-Hermitian bulk instabilities essentially gainless but also actively redirecting arbitrary bulk excitations into a self-stabilizing chiral edge channel.

In Fig.~\ref{fig:fig2} with $\gamma=0.02$ and $\mu=\nu=0.5$ (intermediate coupling regime), we excite a wavepacket at randomly chosen sites $\Psi(\bold r, t=0)$ with $\bold r\in$ bulk.
Initially, the wavepacket spreads non-directionally across the lattice (see Supplementary Sec.~S6 for the early-time regime $t<400$) [Fig.~\ref{fig:fig2}(a), $t=200$], but then it gradually projects onto chiral edge states [$t=400$]. By $t=600$, bulk oscillations diminish relative to the edge excitations, converging towards purely helical propagation.

This self-correcting evolution from arbitrary excitations to chiral edge transport only exists when \emph{both} $\mu$ and $\nu$ couplings are present, as seen by comparing the edge occupation $\rho_{\text{Edge}}(t)\equiv\rho_{\rm Edge}[\Psi(\mathbf r,t)]$ in Figs.~\ref{fig:fig2}(a2) with (b,c). In the latter cases, no distinct edge occupation emerges i.e. $\rho_{\text{Edge}}(t)\approx 0$, similar to the Hermitian ($\gamma=0$) limits (gray curves).

Another striking observation is the feedback-enforced mechanism governing the edge states. In Fig.~\ref{fig:fig2}(a1,a2), the layer-resolved edge wavepacket amplitudes $\Psi(\mathbf{r}\in\mathrm{Edge},t)$ and their edge occupation ratios $\rho_{\text{Edge}}(t)$ oscillate out of step and take turns increasing. Here the ``state'' in the original wording referred to the evolving wavepacket. The layer-resolved oscillations indicate that the wavepacket is repeatedly transferred among the four coupled layers. Because these layers have different combinations of chirality and NHSE direction, the layer that is momentarily favorable for directed amplification enhances its edge component; repeated inter-layer transfer then progressively redirects the total wavepacket toward coherent edge propagation.

\subsection{Effective edge theory for feedback-enforced robustness}
\begin{figure}
  \centering
  \includegraphics[width=0.5\textwidth]{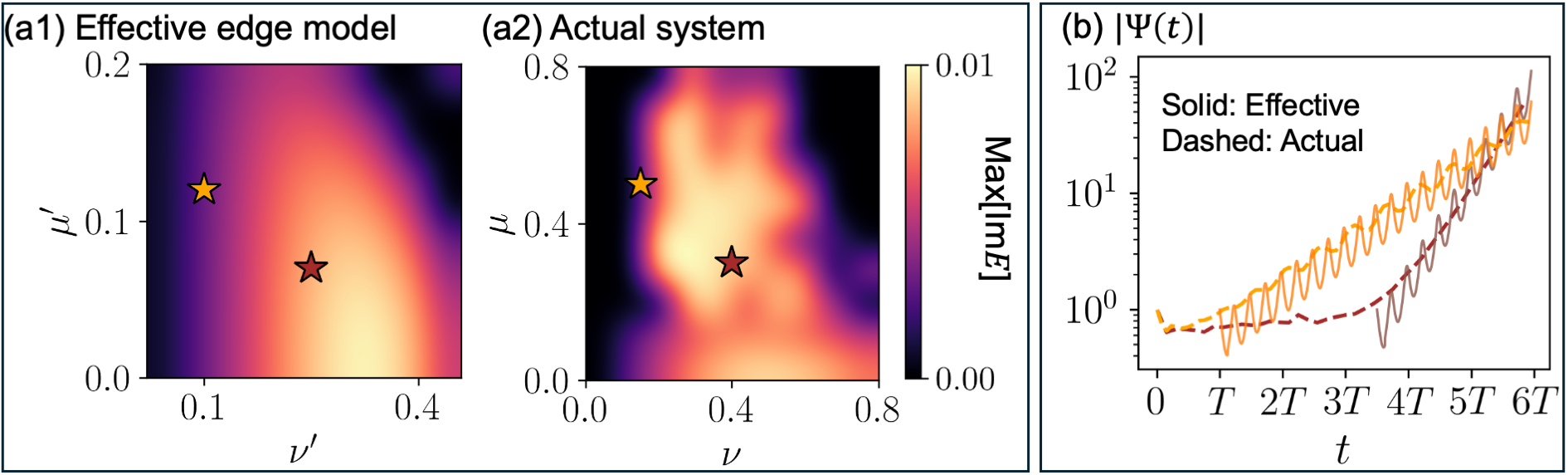}
  \caption{Our effective edge model [Eqs.~\ref{eq:Hstackeff} and \ref{eq:Ueff3}] accurately captures the non-Hermitian growth dynamics of the actual stacked QSH lattice [Eq.~\eqref{eq:biQSH}].
    (a) Good qualitative agreement of the max[Im$E$] (growth rate) between the (a1) effective edge model with $v=0.96,k_y=0.72\pi,\kappa=0.058, T=60$, and the (a2) actual system representing the edge physics from illustrative OBC lattice size $L_x=23, L_y=6$. $\gamma=0.02$. $T=2(L_x+L_y)/v$. $v$ is measured from the actual system.
    (b) Qualitative agreement in the growth of the propagated eigenstates in both the effective edge model (solid curves) and the actual system (dashed curves). Solid curves: $\Psi_{\mathrm{eff}}(\mathbf r,t)=e^{-iH_{\text{stacked-QSH}}t}\phi_{\mathrm{eff}}(\mathbf r)$, with $\phi_{\mathrm{eff}}$ the eigenstate of $H_{\text{stacked-eff}}$ that has the largest imaginary eigenvalue.
  Dashed curves: $\Psi_{\mathrm{edge}}(\mathbf r,t)=e^{-iH_{\text{stacked-QSH}}t}\psi_{\mathrm{edge}}(\mathbf r)$, with $\psi_{\mathrm{edge}}$ the helical edge eigenstate of $H_{\text{stacked-QSH}}$ that has the largest imaginary eigenvalue. Results are shown for two illustrative inter-layer coupling settings [orange and brown, as indicated in (a)]. }
  \label{fig:eff_examples}
\end{figure}

To clarify the physics behind our linear feedback-enforced mechanism, we write down an effective 4-component (layer) evolution operator for the edge states. They are modeled to circumnavigate each layer by alternatively occupying its upper (u) and lower (l) edges for half a period i.e. $T/2=(L_x+L_y)/v$
, with trivial dynamics at the left/right edges (since no NHSE interplay exists). From Fig.~\ref{fig:fig1}(c1), the effective upper/lower edge Hamiltonians are
\begin{equation}
  H_{\text{stacked-u}} =
  \begin{pmatrix}
    k_y v - i\kappa v & \mu' & 0 & \nu' \\
    \mu' & -k_y v - i\kappa v& \nu' & 0 \\
    0 & \nu' & k_y v + i\kappa v& \mu' \\
    \nu' & 0 & \mu' &  i\kappa v -k_y v
  \end{pmatrix},
\end{equation}
\begin{equation}
  H_{\text{stacked-l}} =
  \begin{pmatrix}
    -k_y v+ i\kappa v& \mu' & 0 & \nu' \\
    \mu' & k_y v+ i\kappa v& \nu' & 0 \\
    0 & \nu' & -k_y v - i\kappa v& \mu' \\
    \nu' & 0 & \mu' & k_y v - i\kappa v
  \end{pmatrix}.
\end{equation}
Here, the effective inter-layer couplings $\mu'$ and $\nu'$ are related, but not necessarily equal to the physical couplings $\mu$ and $\nu$. The wavepacket group velocity $v$ directionally couples to $\pm k$ and $\pm i\kappa$, the wave number and non-Hermitian skin depth $\kappa\approx-\log \sqrt{\mid\left(m+\cos k_y-\gamma\right) /\left(m+\cos k_y+\gamma\right)}$.
The overall stroboscopic dynamics is captured by the
Floquet Hamiltonian
\begin{equation}
  H_{\text{stacked-eff}} = -i \frac{\log[U_{\text{stacked}}(T)]}{T},\label{eq:Hstackeff}
\end{equation}
based on the time evolution operator
$U_{\text{stacked}}(t)$
\begin{equation}
  =
  \begin{cases}
    e^{-iH_{\text{stacked-u}}t}, & 0<t<T/2\\
    e^{-iH_{\text{stacked-l}}(t-T/2)}e^{-iH_{\text{stacked-u}}T/2}, & T/2<t<T.
  \end{cases}\label{eq:Ueff3}
\end{equation}
The imaginary part of the eigenvalues of $H_{\text{stacked-eff}}$ [Eq.~\eqref{eq:Hstackeff}] exhibit clear amplification at a dome-shaped intermediate coupling regime [Fig.~\ref{fig:eff_examples}(a1)], qualitatively matching full lattice calculations of $H_\text{stacked-QSH}$ under OBCs [Fig.~\ref{fig:eff_examples}(a2)]. This effective model also reliably reproduces the total state growth, shown in [Fig.~\ref{fig:eff_examples}(b)] for two sets of couplings, one deep in the intermediate regime (brown), the other at its edge (orange). In the former, the dynamics are well-captured only after some time, when the edge modes have grown to be sufficiently well-defined.
Starting from the eigenstate with the largest imaginary part in each Hamiltonian--$\phi_{\text{eff}}$ for $H_{\text{stacked-eff}}$ and the helical-edge eigenstate $\psi_{\text{edge}}$ for $H_{\text{stacked-QSH}}$--evolving $\phi_{\text{eff}}$ with $U_{\text{stacked}}(t)$ soon reproduces the evolution of $\psi_{\text{edge}}$ under the full stacked-QSH Hamiltonian. Evolving an eigenstate of $H_{\text{stacked-eff}}$ under $U_{\text{stacked}}(t)$ closely matches evolving an eigenstate of $H_{\text{stacked-QSH}}$ under itself, i.e., the full stacked-QSH model initialized in a topological edge mode.

\begin{figure}
  \centering
  \includegraphics[width=0.5\textwidth]{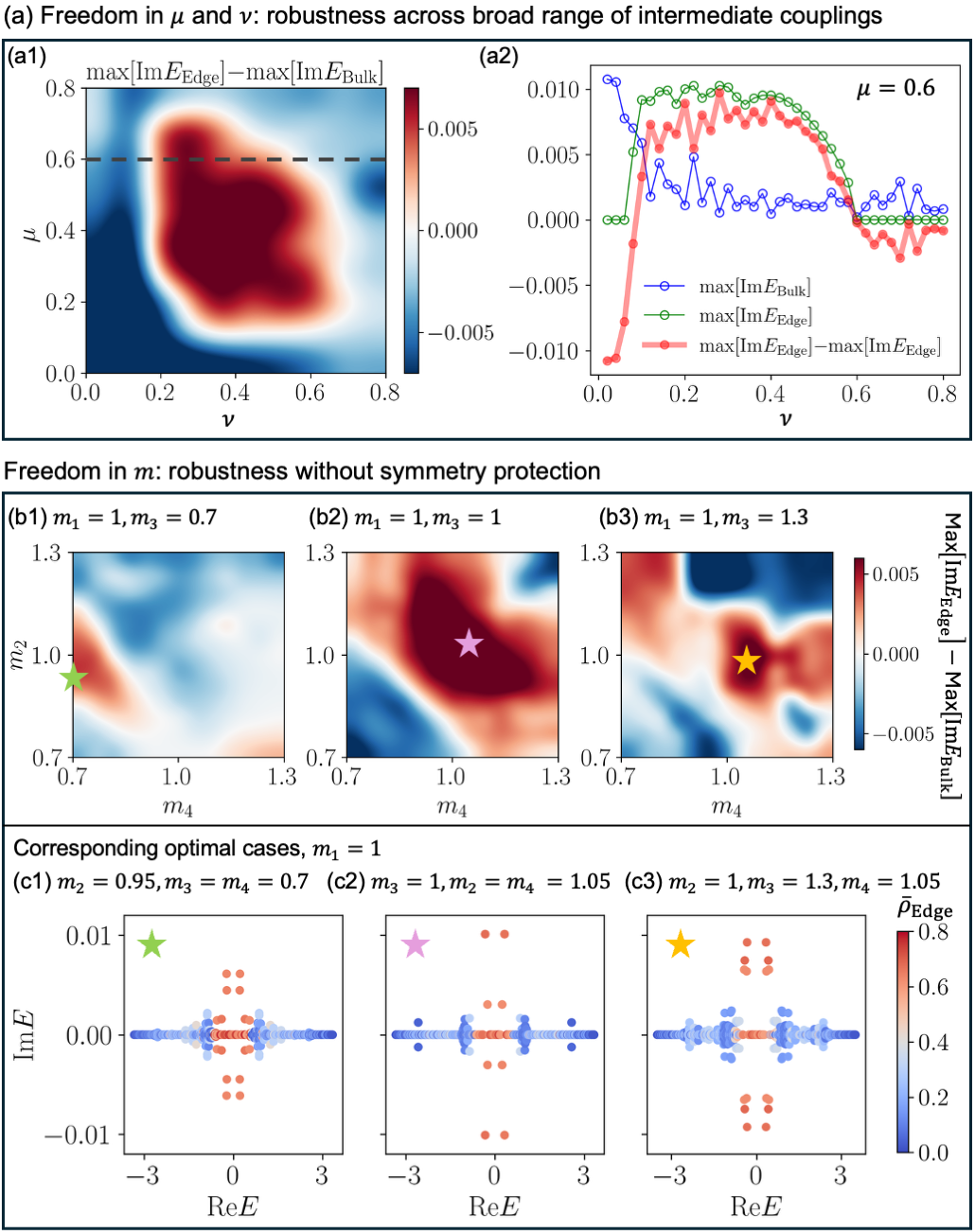}
  \caption{Versatility of feedback-enforced mechanism across wide parameter ranges. (a1) A broad range of $(\mu,\nu)$ gives heightened   max[Im$E_{\text{Edge}}$]$-$max[Im$E_{\text{Bulk}}$]$>0$ (red) of $H_\text{stacked-QSH}$ [Eq.~\eqref{eq:biQSH}] favorable for edge transport. (a2) Cross-section along $\mu=0.6$ (dashed in (a1)) reveals how $\nu$ drives the bulk spectrum essentially real (blue), induces real-to-complex edge state transition (green), and enhances net edge robustness (red) across various strength regimes. (b) Edge transport dominance persists even for unequal $m_1,...,m_4$ across the 4 layers, breaking PT symmetry. (b1-b3) Even for $m_3\neq m_1=1$, broad stability regions exist in parameter space, with optimal cases starred and presented in (c1-c3), each with much greater Im$E$ (amplification) of the in-gap edge states vs. bulk states. Results based on a $L_x\times L_y=40\times 10$ OBC lattice with $\gamma=0.02$. }
  \label{fig:phasediagram}
\end{figure}

\begin{figure}
  \centering
  \includegraphics[width=0.5\textwidth]{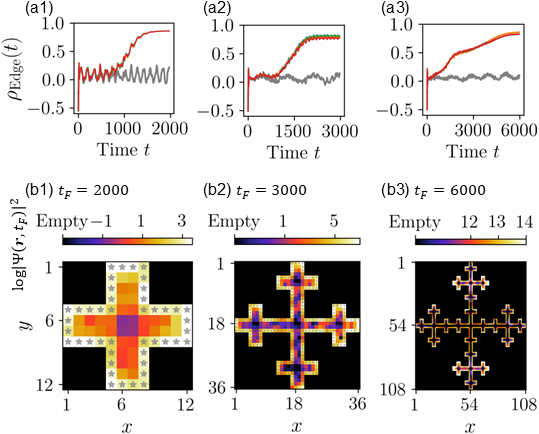}
  \caption{Robustness against boundary irregularities for three fractal generations. (a1)--(a3) Layer-resolved edge occupation ratios $\rho_{\text{Edge}}(t)\equiv\rho_{\text{Edge}}[\Psi(\br,t)]$ [Eq.~\eqref{eq:edge_ratio_general}] for the $12\times12$, $36\times36$, and $108\times108$ lattices, respectively. The four colored curves correspond to the four layers of the non-Hermitian stacked-QSH system, while the gray curve is the Hermitian reference with $\gamma=0$. (b1)--(b3) Final logarithmic densities $\log|\Psi(\br,t_F)|^2$ for the same three lattices at $t_F=2000$, $3000$, and $6000$, respectively. Gray stars mark the boundary sites; the final density is concentrated predominantly on these irregular boundaries. The wavepacket is initialized in the central bulk region and evolves according to $H_{\text{stacked-QSH}}$. The non-Hermitian results use $\gamma=0.02$, $\mu=\nu=0.5$, and $m_1=m_2=m_3=m_4=1$.}
  \label{fig:fractal}
\end{figure}

\subsection{Generality of the feedback-enforced mechanism}
The efficacy of robust edge transport is measured by $\max [ \text{Im} E_{\text{Edge}} ] - \max [ \text{Im} E_{\text{Bulk}} ]$, which sets the inverse time scale for edge dynamics to dominate. This quantity is positive across a wide regime of intermediate $\mu,\nu$ couplings, colored red in Fig.~\ref{fig:phasediagram}(a1). Further examination of the $\mu=0.6$ cross section [Fig.~\ref{fig:phasediagram}(a2)] confirms that at small anti-parallel $\nu$ coupling, increasing $\nu$ rapidly strengthens edge amplification due to increased accessibility to all layers, while suppressing bulk amplification due to NHSE cancellation. However, overly large $\nu$ destroys both the NHSE amplification and topological modes, leading to essentially gainless bulk and edge Im$E$.
Importantly, while the range of desirable $(\mu,\nu)$ couplings is not entirely arbitrary, it exhibits significant versatility.

Moreover, unlike all previously reported symmetry-protected topological edge states~\cite{yuce2018pt,song2020pt,sone2020exceptional,slootman2024breaking,fritzsche2024parity,chaturvedi2025non}, our robust feedback-enforced mechanism does not rely on \emph{any} symmetry, hinging on the distinct ways in which directed NHSE amplification compete among achiral bulk and chiral edge states. \emph{Deliberately} breaking all symmetries through different $m_1,m_2,m_3,m_4$ parameters, which makes all Chern layers inequivalent, there still exist broad red parameter regions where edge states dominate [Figs.~\ref{fig:phasediagram}(b1-b3) and (c1-c3)]. This firmly establishes our mechanism as a versatile and intrinsically symmetry-agnostic approach to enhanced edge robustness.

\subsection{Robustness against extreme boundary irregularities}
To test the robustness of our feedback-enforced edge states against severe boundary irregularities, we apply our protocol to fractal geometries with highly convoluted edges around thin bulk structures. To first establish the robustness of this transport, we have verified that it persists on simpler geometries with sharp, diagonal edges and local defects; see Supplementary Sec.~S11 for additional simulations on geometrically irregular boundaries.
Wavepackets are initialized as bulk states at the fractal center, with propagation confined within fractal boundaries (see Supplementary Sec.~S11 for the self-correcting evolution in the fractal geometry). Despite sharp turns and narrowly-separated edges, the edge states all exhibit very robust propagation with effective gainless bulk diffusion, in the face of varying boundary complexity across different fractal generations [Fig.~\ref{fig:fractal}, gray stars indicate boundaries].
The edge occupation ratio evolution $\rho_{\text{Edge}}(t)$ confirms that after an initial phase of nascent bulk oscillations, which lasts longest in (a1) due to the thicker bulk, the state consistently becomes more and more exclusively located within the edge sites.

\section{Discussion}
In summary, we have identified and explicitly demonstrated a fundamentally new route towards robust helical edge transport by stacking oppositely NHSE-directed  QSH layers at intermediate coupling, overcoming bulk instabilities and symmetry constraints that traditionally limit topological edge states. Though we presented a specific model, the underlying principles of feedback-enforced edge transport require only two ingredients: (i) intermediate inter-layer coupling between two QSH layers and (ii) two QSH layers with different directed non-Hermitian amplification. Our mechanism relies on non-Hermitian feedback-induced redirection and stabilization of bulk and edge states, rather than on topological invariants protected by symmetry.

By removing stringent symmetry and fine-tuning constraints that limits existing topological platforms, our symmetry-agnostic edge transport mechanism substantially expands the technological potential of non-Hermitian topological physics, opening up new avenues in edge-state-based technologies across photonics~\cite{pan2018photonic,xiao2020non,zhu2020photonic,song2020two,ao2020topological,lin2024observation,liu2024observation}, mechanical materials~\cite{brandenbourger2019non,ghatak2020observation,wen2022unidirectional,xiu2023synthetically, li2024observation}, electrical circuits~\cite{hofmann2019chiral,ezawa2019electric,helbig2020generalized,hofmann2020reciprocal,liu2020gain,liu2021non,stegmaier2021topological,zhang2020non,zhang2022observation,shang2022experimental,yuan2023non,zhu2023higher,zhang2024observation,zou2024experimental, zhang2023electrical,stegmaier2024realizing,stegmaier2024topological,shang2024observation,zhang2024observation2,guo2024scale,halder2024circuit,li2025realization}, and programmable universal quantum simulators~\cite{smith2019simulating,gou2020tunable,koh2022simulation,kirmani2022probing,frey2022realization,chertkov2023characterizing,liu2024simulating,yang2023simulating,iqbal2023creation,shen2025observation,koukoutsis2024quantum,koh2024realization, koh2025interacting,zhang2025observation}, many which are extremely tunable. Furthermore, we outline concrete experimental implementations in topolectrical and digital superconducting quantum circuits, together with a practical scan-and-lock protocol that continuously tunes $(\mu,\nu,\gamma)$ to access and maintain the intermediate-coupling window-achieving edge amplification while suppressing bulk excitations (see Supplementary Secs.~S1, S2, S8, and S9).
An enticing next step is to introduce Moiré-modulated inter-layer tunneling~\cite{ohta2012evidence,dai2016twisted,kim2017tunable,naik2018ultraflatbands,cao2018unconventional,cao2018correlated,carr2020electronic,de2022imaging}, where the interplay between twist-controlled minibands and opposing skin effects could unlock an even richer landscape of non-Hermitian topological phenomena.

\section{Methods}

All analytical derivations and numerical procedures used in this work are described in the main text and Supplemental Information.

\section{Resource Availability}
All data supporting the findings of this study are available within the article and its Supplemental Information. Code used to generate the numerical results is available from the authors upon reasonable request.

\clearpage
\newpage
\appendix
\onecolumngrid
\subsection*{\normalsize Supplementary Materials for ``Beyond symmetry protection: Robust feedback-enforced edge states in non-Hermitian stacked quantum spin Hall systems''}

\setcounter{equation}{0}
\setcounter{figure}{0}
\setcounter{table}{0}
\setcounter{section}{0}
\setcounter{page}{1}
\renewcommand{\theequation}{S\arabic{equation}}
\renewcommand{\thefigure}{S\arabic{figure}}
\renewcommand{\thesection}{S\arabic{section}}
\renewcommand{\thepage}{S\arabic{page}}

This PDF contains supporting text, Materials and Methods, supplementary derivations, supplementary figures, and additional implementation notes for the main manuscript.

\setcounter{equation}{0}
\setcounter{figure}{0}
\setcounter{table}{0}
\setcounter{section}{0}
\setcounter{page}{1}
\renewcommand{\theequation}{S\arabic{equation}}
\renewcommand{\thefigure}{S\arabic{figure}}
\renewcommand{\thesection}{S\arabic{section}}
\renewcommand{\thepage}{S\arabic{page}}

This supplement contains the following material in the following sections:
\begin{enumerate}

  \item \textbf{Criterion for a real bulk spectrum in the stacked-QSH model under double PBC}\newline
  We analytically derive the criterion for real bulk spectra under double periodic boundary conditions (PBC), showing how sufficiently strong inter-layer couplings $\mu$ and $\nu$ suppress bulk imaginary parts.

  \item \textbf{Upper bound on the bulk imaginary spectrum under double OBC}\newline
  We derive an explicit upper bound for the residual bulk imaginary spectrum under double open boundary conditions (OBC), including both the Schrieffer--Wolff remainder and exponentially small boundary-overlap corrections.

  \item \textbf{The enhancement of helical states in the stacked-QSH model, even when the constituent Chern layers are fully real}\newline
  We demonstrate numerically that amplified helical edge modes can persist with Im$E_{\text{Edge}}>0$ even when individual constituent Chern layers have nearly real spectra (Im$E_{\text{Chern}}\approx 0$), due to inter-layer coupling.

  \item \textbf{Effective theory on the edge states}\newline
  We develop an effective edge theory for stacked-QSH dynamics, comparing parallel and anti-parallel directional amplification and clarifying when edge spectra remain real or become complex with exponential growth.

  \item \textbf{Bulk and edge point-gap winding and feedback-enforced edge amplification}\newline
  We compute bulk and boundary point-gap winding numbers, showing that trivial bulk winding can coexist with nontrivial edge winding and selective edge amplification in the non-Hermitian stacked-QSH system.

  \item \textbf{Non-directional spreading of the wavepacket in the stacked-QSH model initially within time $t<400$}\newline
  We provide supplementary dynamics in the early-time regime ($t<400$), where a bulk-initialized wavepacket spreads isotropically before feedback-driven edge localization dominates.

  \item \textbf{Reality of edge spectra under $x$- and $y$-OBC in NH Chern and NH QSH insulators}\newline
  We provide a real-space pseudo-Hermitian/$\mathcal{PT}$-symmetric argument explaining why edge spectra remain essentially real under open boundaries in both spatial directions.

  \item \textbf{Possible experimental realization in a classical topoelectrical circuit}\newline
  We present a circuit-based implementation using INIC and passive components, and map effective model parameters to experimentally accessible capacitances/inductances and readout protocols.

  \item \textbf{Possible quantum-circuit realization on a digital superconducting processor}\newline
  We outline a gate-level realization that combines compiled chiral dynamics with ancilla-assisted nonunitary pumping, including Trotterization, inter-layer coupling synthesis, and PT-threshold diagnostics.

  \item \textbf{Toward feedback-enforced edge-selected lasing}\newline
  We discuss how the stacked non-Hermitian QSH architecture can realize edge-selected lasing under uniform pumping, guided by the same coupling criterion that keeps the bulk effectively real.

  \item \textbf{The self-correcting evolution in the fractal geometry}\newline
  We provide additional fractal-geometry snapshots showing robust self-correcting evolution from bulk spreading toward irregular-boundary localization at long times.
  
  \end{enumerate}

 \section{Criterion for a real bulk spectrum in the stacked-QSH model under double PBC}

In the main text we showed that the double-PBC bulk spectrum of the stacked-QSH model can be made real by tuning the inter-layer couplings $\mu$ and $\nu$ to be sufficiently large, i.e.\ $\mu\nu\ge\gamma$, as indicated by the dashed hyperbola in Fig.~1(c3). Here we provide a detailed analytical derivation of this criterion and discuss its generality.

\vspace{2em}
\emph{(1) Identifying the most non-Hermitian Bloch point}

For the parameters used in Fig.~1, the Bloch momentum that is most susceptible to complexification is
\begin{equation}
\mathbf k^\star=(\pi/2,0),
\end{equation}
where the Hermitian part of the single-layer Chern Hamiltonian is minimal while the non-Hermitian term $i\gamma\sigma_y$ is maximal. With $m=1$, the single-layer Bloch Hamiltonian
\begin{equation}
H_{\text{Ch}}(\mathbf k, m, \gamma)
= \bigl(m + \cos k_x + \cos k_y\bigr)\sigma_x
+ \bigl(i\gamma + \sin k_x\bigr)\sigma_y
+ \sin k_y \sigma_z
\label{eq:single_chern_full_reply}
\end{equation}
reduces at $\mathbf k^\star$ to
\begin{equation}
H_{\text{Ch}}\bigl(\mathbf k^\star\bigr)=
2\sigma_x+\bigl(1+i\gamma\bigr)\sigma_y .
\label{eq:single_chern_kstar_reply}
\end{equation}
The eigenvalues of Eq.~\eqref{eq:single_chern_kstar_reply} can be written as
\begin{equation}
\varepsilon_\pm = E_\pm^{\rm (Re)} + i\gamma,
\end{equation}
so that the two eigenstates are separated by a purely \emph{real} energy gap
\begin{equation}
\Delta E = E_+^{\rm (Re)} - E_-^{\rm (Re)} .
\end{equation}
In the parameter regime of Fig.~1, this gap $\Delta E$ is of order unity and fixes the natural energy scale near $\mathbf k^\star$.

Thus, at $\mathbf k^\star$ the spin sector can be viewed as a pair of states with equal gain $+i\gamma$ but separated by a real gap $\Delta E$.

\emph{(2) Four-layer structure and block form}

We now consider the full stacked-QSH model at $\mathbf k^\star$, consisting of four Chern layers with inter-layer couplings $\mu$ and $\nu$ arranged in the pattern described in the main text. Diagonalizing the spin sector of each layer and grouping the resulting states by their gain/loss, we reorder the basis as
\begin{equation}
\bigl\{|1\rangle_{+\gamma},|4\rangle_{+\gamma},|2\rangle_{-\gamma},|3\rangle_{-\gamma}\bigr\},
\end{equation}
where the subscript $\pm\gamma$ indicates whether the corresponding single-layer state carries net gain ($+\gamma$) or loss ($-\gamma$) at $\mathbf k^\star$.

In this reordered basis, the $8\times 8$ Bloch Hamiltonian takes a Bogoliubov--de~Gennes-like block form
\begin{equation}
\mathcal H(\mathbf k^\star)=
\begin{pmatrix}
+i\gamma\openone_2 + H_{\rm gain}^{\rm (Re)} & V \\
V^\dagger & -i\gamma\openone_2 + H_{\rm loss}^{\rm (Re)}
\end{pmatrix},
\label{eq:bdg_block_general_reply}
\end{equation}
where $H_{\rm gain}^{\rm (Re)}$ and $H_{\rm loss}^{\rm (Re)}$ are Hermitian $2\times2$ matrices encoding the real parts of the energies in the gain and loss sectors, respectively, and are separated by a typical real gap of order $\Delta E$. The off-diagonal block
$V$ arises from inter-layer couplings. For the stacked-QSH geometry used in Fig.~1, one obtains
\begin{equation}
V = \mu\,\openone_2 + \nu\,\sigma_x .
\label{eq:V_specific_reply}
\end{equation}
The upper-left block of Eq.~\eqref{eq:bdg_block_general_reply} describes two gain-dominated modes (with $+i\gamma$), while the lower-right block describes two loss-dominated modes (with $-i\gamma$). The real parts of these blocks differ by the energy gap $\Delta E$.

\emph{(3) Schrieffer-Wolff reduction}

We now perform a second-order Schrieffer--Wolff (SW) transformation to obtain an effective Hamiltonian within the gain sector, by integrating out the loss sector that is separated by the real gap $\Delta E$. For a block-partitioned Hamiltonian of the form
\begin{equation}
\mathcal H =
\begin{pmatrix}
A & V \\ V^\dagger & D
\end{pmatrix},
\end{equation}
standard SW theory yields, to second order in $V$ and for energies within the spectral window of $A$,
\begin{equation}
H_{\rm eff} \approx A - V D^{-1} V^\dagger .
\label{eq:sw_general_reply}
\end{equation}
In our case, we identify
\begin{equation}
A = +i\gamma\openone_2 + H_{\rm gain}^{\rm (Re)}, \qquad
D = -i\gamma\openone_2 + H_{\rm loss}^{\rm (Re)} .
\end{equation}
The Hermitian parts of $A$ and $D$ are separated by the real gap $\Delta E$, so to leading order we may approximate
\begin{equation}
D^{-1} \simeq \frac{1}{\Delta E}\,\openone_2 + \mathcal O\bigl(1/\Delta E^2\bigr).
\end{equation}
Substituting into Eq.~\eqref{eq:sw_general_reply} gives
\begin{equation}
H_{\rm eff}
\simeq +i\gamma\openone_2 + H_{\rm gain}^{\rm (Re)}
- \frac{1}{\Delta E} V V^\dagger
+ \mathcal O\Bigl(\frac{\|V\|^3}{\Delta E^2}\Bigr).
\label{eq:Heff_before_detail_reply}
\end{equation}

Using the explicit form of $V$ in Eq.~\eqref{eq:V_specific_reply},
\begin{equation}
V V^\dagger
= (\mu\openone_2 + \nu\sigma_x)(\mu\openone_2 + \nu\sigma_x)
= (\mu^2 + \nu^2)\openone_2 + 2\mu\nu\,\sigma_x .
\end{equation}
The contribution proportional to $\openone_2$ merely shifts the real part of the spectrum uniformly and does not affect whether the eigenvalues are real or complex. Thus we isolate the part that controls the \emph{relative} splitting and complexification:
\begin{equation}
H_{\rm eff}
\simeq +i\gamma\openone_2
- \frac{2\mu\nu}{\Delta E}\,\sigma_x
+ \bigl[\text{overall real shift}\bigr]
+ \mathcal O(\mu^2)+\mathcal O(\nu^2).
\label{eq:Heff_mu_nu_reply}
\end{equation}
Dropping the overall real shift and higher-order terms, Eq.~\eqref{eq:Heff_mu_nu_reply} is equivalent to the $\mathcal PT$-symmetric dimer
\begin{equation}
H_{\mathcal PT} =
\begin{pmatrix}
 i\gamma & J_{\rm eff}\\
 J_{\rm eff} & -i\gamma
\end{pmatrix},
\qquad
J_{\rm eff} = \frac{2\mu\nu}{\Delta E}.
\label{eq:PT_dimer_reply}
\end{equation}

\emph{(4) $\mathcal PT$-symmetry criterion and the product $\mu\nu\ge\gamma$}

The eigenvalues of the dimer in Eq.~\eqref{eq:PT_dimer_reply} are
\begin{equation}
E_\pm = \pm\sqrt{J_{\rm eff}^2 - \gamma^2}.
\end{equation}
Thus the spectrum of $H_{\mathcal PT}$ is purely real when
\begin{equation}
|J_{\rm eff}| \ge \gamma.
\label{eq:PT_condition_general_reply}
\end{equation}
Substituting $J_{\rm eff}=2\mu\nu/\Delta E$ gives
\begin{equation}
\left|\frac{2\mu\nu}{\Delta E}\right| \ge \gamma.
\label{eq:criterion_general_with_gap_reply}
\end{equation}
In the parameter regime of Fig.~1, $\Delta E$ is a fixed real number of order unity (set by the single-layer dispersion at $\mathbf k^\star$ and the chosen units). Absorbing the factor $2/\Delta E$ into the definition of $\mu\nu$ leads to the simple and physically transparent condition
\begin{equation}
\boxed{\mu \nu \geq \gamma},
\label{eq:criterion_simple_reply}
\end{equation}
which is the criterion quoted in the main text and used to plot the dashed hyperbola in Fig.~1(c3). The essential point is that the \emph{product} $\mu\nu$ appears because the effective coupling between gain and loss sectors is generated by a \emph{second-order} virtual process proportional to $V V^\dagger$, rather than by a single bare hopping amplitude.

\emph{(5) Generality for four-layer Chern stacks}

The above derivation relies only on three structural ingredients that are generic for stacked non-Hermitian Chern systems:
(i) around the most non-Hermitian Bloch point $\mathbf k^\star$, the spectrum can be approximately decomposed into a gain-dominated sector and a loss-dominated sector, separated by a real gap $\Delta E(\mathbf k^\star)$;
(ii) inter-layer couplings generate an off-diagonal block $V$ that connects these sectors;
(iii) the shortest virtual process that couples gain and loss sectors involves a product of hoppings along a path that traverses the stack.

In a generic four-layer Chern stack, the SW reduction around $\mathbf k^\star$ again yields an effective $\mathcal PT$ dimer of the form
\begin{equation}
H_{\mathcal PT}^{\rm(gen)} =
\begin{pmatrix}
 i\gamma_{\rm eff} & J_{\rm eff}(\mathbf k^\star)\\
 J_{\rm eff}(\mathbf k^\star) & -i\gamma_{\rm eff}
\end{pmatrix},
\end{equation}
where $J_{\rm eff}(\mathbf k^\star)$ is proportional to a homogeneous polynomial of the inter-layer couplings of degree equal to the length of the shortest path connecting the gain and loss sectors. In our stacked-QSH geometry this path has length two, so
\begin{equation}
J_{\rm eff}(\mathbf k^\star)\propto \frac{\mu\nu}{\Delta E(\mathbf k^\star)}.
\end{equation}
For other stacking patterns the explicit form of $J_{\rm eff}$ may differ (e.g.\ involving $\mu^2$, $\nu^2$ or more couplings), but the \emph{structure} of the criterion
\begin{equation}
|J_{\rm eff}(\mathbf k^\star)| \gtrsim \gamma_{\rm eff}
\end{equation}
remains the same: the onset of a purely real bulk spectrum is controlled by a balance-of-scales condition between an effective coherent coupling $J_{\rm eff}$ and the non-Hermitian strength $\gamma_{\rm eff}$. In our concrete model, this balance reduces to the simple hyperbolic condition $\mu\nu\ge\gamma$.

\emph{(6) Physical meaning: bulk suppression versus edge enhancement}

Finally, we emphasize the physical content of Eq.~\eqref{eq:criterion_simple_reply}. The derivation above is performed under \emph{double PBC} and therefore characterizes the \emph{bulk} Bloch spectrum. When $\mu\nu < \gamma$, the condition \eqref{eq:PT_condition_general_reply} is violated at $\mathbf k^\star$, and the bulk spectrum is complex: bulk modes carry net gain or loss and can grow or decay. When $\mu\nu \ge \gamma$, the bulk spectrum becomes (almost) purely real, putting the bulk into a $\mathcal PT$-unbroken regime where excitations are neither exponentially amplified nor depleted, leading to the strong suppression of bulk intensity observed in Fig.~1.

In contrast, open boundaries break the exact cancellation mechanism that underlies Eq.~\eqref{eq:criterion_simple_reply}. The feedback loop responsible for balancing gain and loss is interrupted, so edge-localized modes retain residual imaginary parts even when the bulk Bloch spectrum is real. As a result, excitations preferentially accumulate and are amplified at the edges, while remaining suppressed in the interior. This provides an analytical understanding of why bulk excitations are strongly suppressed inside the bulk but enhanced around the edge, going beyond purely numerical observations.

\section{Upper bound on the bulk imaginary spectrum under double OBC}
\label{sec:OBC-bound}

{Despite the previous section showing that, in the Bloch (double-PBC) analysis, a \emph{purely real} bulk spectrum is ensured once $\mu\nu\ge \gamma$, under double OBC we nevertheless observe a \emph{nonzero} bulk $\operatorname{Im}E$.} Throughout the main text we therefore describe the double-OBC bulk spectrum as ``essentially real.''

Here we derive an explicit upper bound on this residual $\operatorname{Im}E$ and identify its two sources: (i) a size-independent Schrieffer-Wolff (SW) remainder arising from the imperfect cancellation between gain/loss blocks after block elimination, and (ii) an exponentially small boundary self-energy from the overlap of left/right skin tails under OBC. Both contributions are parametrically small; in particular the total bound lies \emph{far below} the single-layer Chern model’s imaginary scale $|\operatorname{Im}E|=\gamma$.

As around the hyperbolic boundary discussed previously, the most dangerous quasi-momentum for our parameters is $\bf k^\star=(\pi/2,0)$. At $m=1$ the single-layer Chern block has eigenvalues $\varepsilon_\pm=\pm 2+i\gamma$, so the two gain/loss blocks are split in the \emph{real} part by $\Delta E_\ast=2$. Reordering the four Chern layers as $\{|1\rangle_{+\gamma},|4\rangle_{+\gamma},|2\rangle_{-\gamma},|3\rangle_{-\gamma}\}$, the $8\times8$ Hamiltonian near $\bf k^\star$ takes the Bogoliubov-de Gennes form
\begin{equation}
\mathcal H=
\begin{pmatrix}
+i\gamma\openone_2 & V\\
V^\dagger & -i\gamma\openone_2
\end{pmatrix}
\oplus (\text{remote real bands}),
\qquad
V=\mu\openone_2+\nu\sigma_x .
\label{eq:OBC-block}
\end{equation}
Performing an SW elimination of the lower block and collecting the boundary-induced self-energy into the upper block yields the effective $2\times2$ problem
\begin{equation}
H_{\rm eff}
= i\gamma\openone_2
-\frac{1}{\Delta E_\ast}V V^\dagger
+\underbrace{\mathcal R_{\rm SW}}_{\text{SW remainder}}
+\underbrace{R(L)}_{\text{boundary}} .
\label{eq:Heff-OBC}
\end{equation}
Away from exceptional points, the anti-Hermitian SW remainder obeys the norm estimate $\|\operatorname{Im}\mathcal R_{\rm SW}\|_2\lesssim \gamma\|V\|_2^2/\Delta E_\ast^2$, while the boundary self-energy obeys $\|R(L)\|_2\le Ce^{-L/\xi}$, with $\xi$ the skin depth set by the nonreciprocity and $C=O(1)$ depending on termination and target energy.

Taking spectral norms, and using $\|V\|_2=\sqrt{\mu^2+\nu^2}$, we arrive at the bulk upper bound
\begin{equation}
\boxed{
\max_{\rm bulk}\bigl|\operatorname{Im}E\bigr|
\lesssim
\frac{\gamma(\mu^2+\nu^2)}{\Delta E_\ast^2}
+
Ce^{-L/\xi(\mu,\nu,\gamma)}
} .
\label{eq:OBC-bound}
\end{equation}
The first term is the size-independent SW remainder; the second is the finite-size (non-Bloch) correction from skin-tail overlap. {Both are strongly suppressed in the bulk-real dome ($\mu\nu\gtrsim\gamma$).}

\paragraph*{On the size of $C$ ($C=O(1)$).}
The boundary term arises as an end-to-end self-energy
\begin{equation}
R(L;E_0)=B_L^\dagger G_{1L}(E_0)B_R+\text{h.c.},
\label{eq:C-endtoend}
\end{equation}
with $B_{L,R}$ the edge-bulk couplings (hard walls give $\|B_{L,R}\|=O(1)$) and $G_{1L}(E_0)=(H_{\rm bulk}-E_0)^{-1}$ the resolvent from the leftmost to the rightmost unit at a target bulk energy $E_0$. Away from exceptional points and band closings, standard resolvent/transfer-matrix estimates give
\begin{equation}
\|G_{1L}(E_0)\|\le c(E_0)e^{-L/\xi},
\qquad
c(E_0)\lesssim\frac{\kappa(X)}{\Delta_{\rm loc}},
\label{eq:G-bound}
\end{equation}
where $\kappa(X)=\|X\|\|X^{-1}\|$ is the eigenbasis condition number of $H_{\rm bulk}=X\Lambda X^{-1}$ and $\Delta_{\rm loc}=\operatorname{dist}(E_0,\sigma(H_{\rm bulk}))$. Hence
\begin{equation}
\|R(L;E_0)\|
\le
2\|B_L\|\|B_R\|\frac{\kappa(X)}{\Delta_{\rm loc}}e^{-L/\xi}
\equiv Ce^{-L/\xi}.
\label{eq:R-bound}
\end{equation}
In our dimensionless units (nearest-neighbour scale $t\sim 1$, $\mu,\nu=O(1)$, $\gamma\ll 1$), hard-wall terminations give $\|B_{L,R}\|=O(1)$; inside the bulk-real dome we are far from EPs and band closings, so $\kappa(X)=O(1\text{-}10)$ and $\Delta_{\rm loc}=O(1)$. Therefore
\begin{equation}
C\lesssim2\|B_L\|\|B_R\|\kappa(X)/\Delta_{\rm loc}=O(1)
\quad(\text{at most }\sim10),
\label{eq:C-O1}
\end{equation}
independent of $L$ and without exponential growth. Only near EPs, near band closings, or under special surface resonances can $C$ become anomalously large; these regimes are excluded by our parameter choices.

\paragraph*{Magnitude approximation for our parameters ($L=40$).}
For $m=1$ we take $\Delta E_\ast=2$, so the size-independent term simplifies to $\tfrac{\gamma}{4}(\mu^2+\nu^2)$. Representative values used in our numerics give
\begin{equation}
\gamma=0.02,\mu=\nu=0.5 \Rightarrow \frac{\gamma}{4}(\mu^2+\nu^2)=2.5\times10^{-3}.
\end{equation}
For the finite-size correction at $L=40$ we take a conservative $\xi\simeq 3\text{-}5$ and $C=1\text{-}5$:
\begin{equation}
C e^{-L/\xi}
= C e^{-40/\xi}
\in \bigl[1.6\times10^{-6},1.7\times10^{-3}\bigr].
\end{equation}
Hence, under double OBC the residual bulk $|\operatorname{Im}E|$ is at the $10^{-3}$ level from the SW remainder plus an exponentially small $L$-dependent tail--\emph{much smaller} than the single-layer Chern model’s $|\operatorname{Im}E|=\gamma=2\times 10^{-2}$, which justifies our wording ``essentially real.''

\color{black}

\section{The enhancement of helical states in the stacked-QSH model, even when the constituent
Chern layers are fully real} 

\begin{figure}
  \centering
  \subfigure[]{\includegraphics[width=0.3\textwidth]{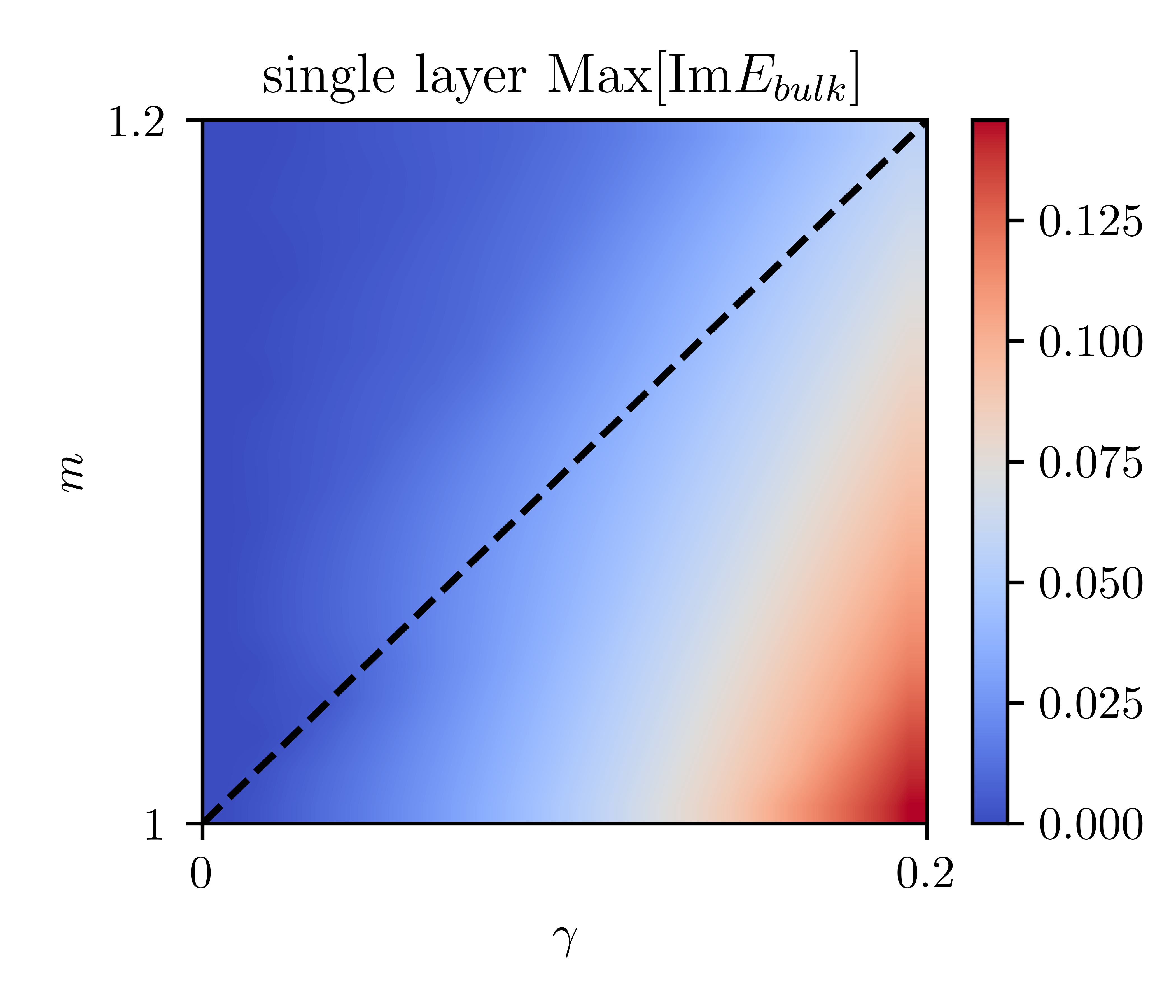}}
  \subfigure[]{\includegraphics[width=0.3\textwidth]{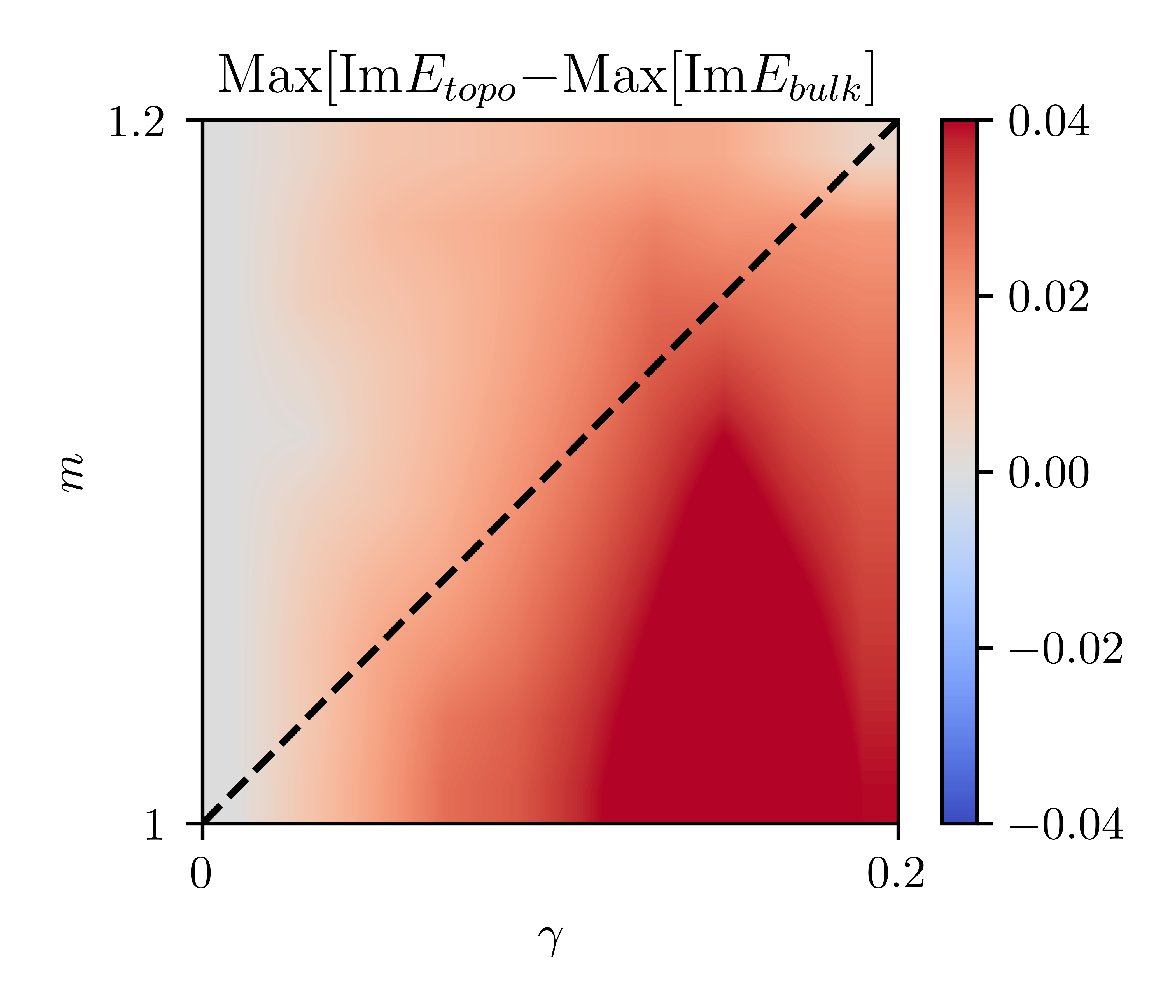}}
  \caption{The presence of edge modes with amplified imaginary energies results from the interlayer coupling between Hermitian-like layers. (a) The maximum imaginary eigenenergy of a single-layer Chern insulator, with the blue region indicating predominantly real eigenenergies ($\text{Im}E_{\text{Chern}}\approx 0$). (b) The difference between the maximum imaginary eigenenergy at the edges and in the bulk of the stacked-QSH model with $L_x=20,L_y=6,\mu=\nu=0.5$. The dominant red regions in (b) highlight that the stacked-QSH model significantly amplifies topological edge states, yielding larger imaginary eigenenergies at the edges compared to the bulk, even when the constituent Chern layers individually exhibit negligible imaginary components. The black dashed lines separate the $m-1>\gamma$ and $m-1<\gamma$ regimes. }
  \label{fig:gamma-m-phase}
\end{figure}

In the main text, we demonstrated that the stacked-QSH model can host helical edge states characterized by complex eigenenergies with an imaginary part exceeding that of the bulk eigenstates, i.e., Im$E_{\text{Edge}}>0$ and Im$E_{\text{Bulk}}=0$. This phenomenon is realized by stacking two quantum spin Hall (QSH) layers at intermediate coupling strengths, each composed of constituent non-Hermitian Chern layers with intrinsically complex eigenenergies. Specifically, each Chern layer exhibits Im$E_{\text{Chern}}\neq0$ when the condition $m-1<\gamma$ is satisfied.

Remarkably, it is also possible to achieve helical edge states with Im$E_{\text{Edge}}>\text{Im}E_{\text{Bulk}}$ even when stacking two QSH layers whose constituent Chern layers individually possess fully real eigenenergies, i.e., Im$E_{\text{Chern}}=0$. This scenario emerges under the condition $m-1>\gamma$.

In Fig.~\ref{fig:gamma-m-phase}, we further illustrate that the presence of edge modes with amplified imaginary energies arises specifically from interlayer coupling between Hermitian-like layers. Fig.~\ref{fig:gamma-m-phase}(a) shows the maximum imaginary eigenenergy of a single-layer Chern insulator, where the black dashed line approximately separates regions of predominantly real (blue) and complex (red) eigenenergies. In contrast, Fig.~\ref{fig:gamma-m-phase}(b) displays the difference between the maximum imaginary eigenenergy at the edges and in the bulk of the stacked-QSH model with parameters $L_x=20$, $L_y=6$, and $\mu=\nu=0.5$. Notably, the red regions dominate nearly the entire parameter space in Fig.~\ref{fig:gamma-m-phase}(b), indicating significant amplification of edge states. Moreover, interlayer coupling induces substantial edge amplification even in parameter regions where individual layers have predominantly real eigenenergies (blue region in Fig.~\ref{fig:gamma-m-phase}(a)).

\section{Effective theory on the edge states}

In the main text, we have claimed that stacking two QSH insulators with intermediate coupling strength—neither too strong (which would cause cancellation of helical states and yield a trivial phase) nor too weak (failing to establish effective four-layer connectivity)—ensures complex edge Im$E$ and vanishing bulk Im$E$. For the vanishing bulk Im$E$, we already discussed in the main text Fig.~1(c2). To support our claim on topological states, we provide an effective theory for the edge states below. 

\begin{figure}
  \centering
  \includegraphics[width=0.6\textwidth]{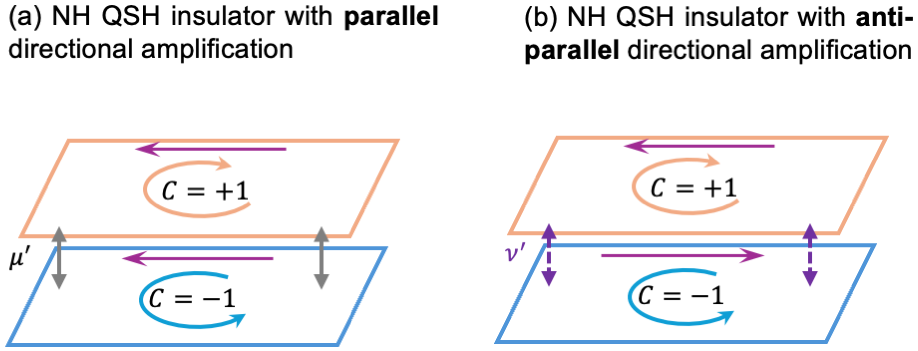}
  \caption{Schematic of the effective theory for the edge states. (a) In a single QSH layer with parallel directional amplification, the edge state eigenenergies remain real regardless of the coupling strength $\mu'$. (b) In a single QSH layer with anti-parallel directional amplification, the edge state eigenenergies become complex. }
  \label{fig:eff_schematic}
\end{figure}

\begin{itemize}
  \item In the first subsection, Sect.~\ref{sec:eff1}, we show that in a single QSH layer [Fig.~\ref{fig:eff_schematic}(a), i.e., the model in Fig.~1(b1)] with \textbf{parallel directional amplification}, i.e., the skin effect direction is the same, the edge state eigenenergies remain real regardless of the coupling strength $\mu'$.
  \item In the second subsection, Sect.~\ref{sec:eff2}, we show that the edge state eigenenergies become complex when the \textbf{directional amplification is anti-parallel}, i.e., the skin effect directions are opposite [Fig.~\ref{fig:eff_schematic}(b)], corresponding to the case where the two layers have opposite skin effect directions.
\end{itemize}

\subsection{Real spectrum in one QSH layer with parallel directional amplification}\label{sec:eff1}

Now that we only consider the edge spectrum, we can retrieve only the edge dynamics signature from the QSH model in Fig.~\ref{fig:eff_schematic}(a). The edge states evolving on the upper edge are described by the effective Hamiltonian as 
\begin{equation}
  H_{u} = \begin{pmatrix}
    k_y v - i\kappa v & \mu' \\
  \mu' & -k_y v - i\kappa v
  \end{pmatrix},
\end{equation} 
where $\kappa$ is the skin depth of the edge states. It is noted that we also need to consider the lower edge, and the effective Hamiltonian for the lower edge is given by
\begin{equation}
  H_{l} = \begin{pmatrix}
    -k_y v + i\kappa v& \mu' \\
  \mu' & k_y v + i\kappa v
  \end{pmatrix}.
\end{equation}

Let the dynamics of an initial wavepacket $\Phi(t=0)=\left(\Phi^\uparrow,\Phi^\downarrow\right)^T$ in one period 
\begin{equation}
  T=2L_x/v,
\end{equation}
which can be expressed as $U_T \Phi(t=0)$. $L_x$ is the length of the system in the $x$ direction, i.e., the directional amplification direction. It is noted that, here, we neglect the width of the system in the $y$ direction, which does not affect the amplification dynamics, as the skin effect is only along the $x$ direction.

The time evolution operator $U_T$ can be expressed as
\begin{equation}
  U(t) = 
  \begin{cases}
    \exp(-iH_{u}t), & 0<t<T/2\\
    \exp(-iH_{l}(t-T/2))\exp(-iH_{u}T/2), & T/2<t<T.
  \end{cases}\label{eq:Ueff1}
\end{equation}
In this way, we can get the wavepacket after one period $T$ as $\Phi(t=T) = U(t=T) \Phi(t=0)$. If we denote $\mu'^2+k_y^2=\omega^2$, the final state $\Phi(t=T)$ can be expressed as 
\begin{equation}
  \begin{aligned}
  \phi_1(T)&=\Phi^\uparrow\frac{\mu'^2\cos(\omega T) + k_y^2}{\omega^2}  + \Phi^\downarrow\left[\frac{\mu'\left(1-e^{-i\omega T}\right)}{2\omega^2}\Bigl(k(1-e^{i\omega T})-\omega(1+e^{i \omega T})\Bigr)\right],\\
  \phi_2(T)&=
    \Phi^\uparrow\left[\frac{\mu'\left(1-e^{-i\omega T}\right)}{2\omega^2}\Bigl(k(-1+e^{i\omega T})-\omega(1+e^{i \omega T})\Bigr)\right]+\Phi^\downarrow
    \frac{\mu'^2\cos(\omega T) + k_y^2}{\omega^2}
  \end{aligned}
\end{equation}
In this way, the effective Hamiltonian for the edge states can be obtained as
\begin{equation}
  H_{\text{eff}} = \begin{pmatrix}
    \frac{\mu'^2\cos(\omega T) + k_y^2}{\omega^2} & \left[\frac{\mu'\left(1-e^{-i\omega T}\right)}{2\omega^2}\Bigl(k_y(1-e^{i\omega T})-\omega(1+e^{i \omega T})\Bigr)\right]  \\
    \left[\frac{\mu'\left(1-e^{-i\omega T}\right)}{2\omega^2}\Bigl(k_y(-1+e^{i\omega T})-\omega(1+e^{i \omega T})\Bigr)\right]  & \frac{\mu'^2\cos(\omega T) + k_y^2}{\omega^2}
  \end{pmatrix}.
\end{equation}
The eigenenergies of $H_{\text{eff}}$ are 
\begin{equation}
  \begin{aligned}
  E_{\text{eff}}& = \frac{\mu'^2\cos(\omega T) + k_y^2}{\omega^2} \pm \frac{\mu'}{2\omega^2}\sqrt{\left[\omega(1+e^{i \omega T})\left(1-e^{-i\omega T}\right)\right]^2-\left[k_y(1-e^{i\omega T})\left(1-e^{-i\omega T}\right)\right]^2}.\\
  &= \frac{\mu'^2\cos(\omega T) + k_y^2}{\omega^2} \pm \frac{\mu'}{2\omega^2}\sqrt{4\omega^2\sin^2(\omega T)+4k_y^2\sin^2(\omega T)}\\
  &= \frac{\mu'^2\cos(\omega T) + k_y^2}{\omega^2} \pm \frac{\mu'\sin(\omega T)}{\omega^2}\sqrt{\omega^2+k_y^2}.
  \end{aligned}\label{eq:Eeff1}
\end{equation}
It is noted that both eigenenergies are real, then the edge states are not amplified. This explains why we see purely real edge spectrum in Fig.~1(b3) in the main text.

\subsubsection{Example}

\begin{figure}[H]
  \centering
  \includegraphics[width=0.8\textwidth]{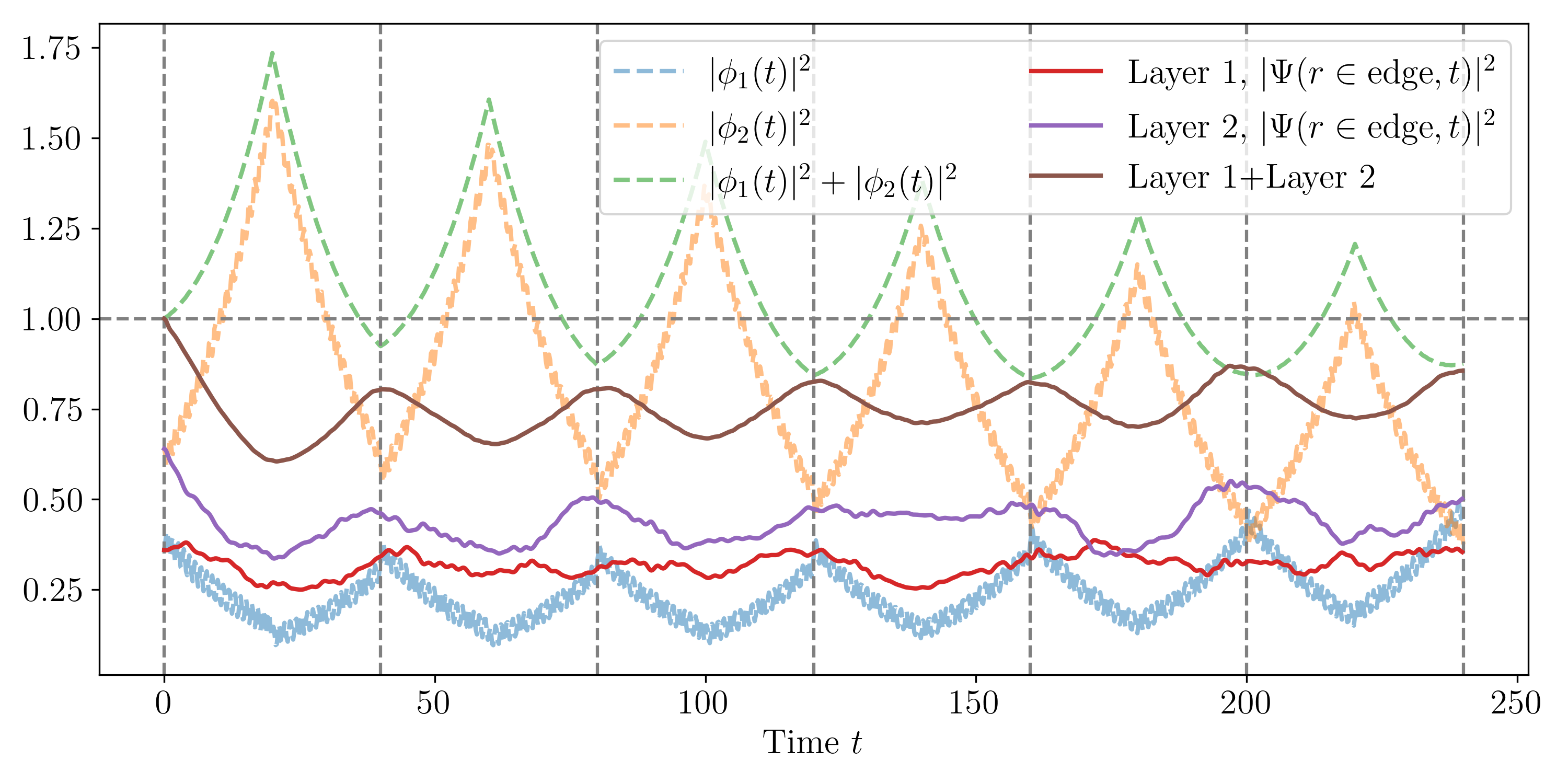}
  \caption{Time evolution of the edge states in an effective model [Eq.~\eqref{eq:Ueff1}] of a single QSH layer with the same amplification directions. The initial state is $\Phi(t=0) = (0.6, 0.8)^T$, with parameters $\mu'=0.2$, $k_y v=3, T=40$, and $\kappa v=0.024, v=0.96$. The wavepacket exhibits periodic motion over six periods without any amplification, indicating the absence of exponential growth in the edge states, which agrees that 1) $E_{\text{eff}}$ [Eq.~\eqref{eq:Eeff1}] are real, and 2) the edge spectrum in Fig.~1(b3) is purely real. Parameters in the actual system are $\gamma=0.02$, $\mu=0.2$, $L_x=14$, and $L_y=4$. }
  \label{fig:eff1}
\end{figure}

To verify the analytical results in $E_{\text{eff}}$, we conduct numerical simulations in the effective model in a single QSH layer with the same skin effect. The initial state is set as $\Phi(t=0) = (0.6, 0.8)^T$, with parameters $\mu'=0.2$, $k_y v=3$, and $\kappa=0.1$. The time evolution $\Phi(t)=(\phi_1(t),\phi_2(t))^T=U(t)\Phi(t=0)$ in Eq.~\eqref{eq:Ueff1} of the wavepacket over six periods is shown in Fig.~\ref{fig:eff1}. It is observed that the wavepacket undergoes periodic motion without amplification. This confirms that in a single QSH layer with the same skin effect, the edge states remain bounded and do not exhibit exponential growth.

\subsection{Complex spectrum in one QSH layer with anti-parallel directional amplification}\label{sec:eff2}
\subsubsection{Analytical derivation}

Now we want to consider the effective model on topological states in Fig.~\ref{fig:eff_schematic}(b), where $\nu'$ connects layers with opposite skin effect. Follow the same procedure as in the previous subsection, and the effective Hamiltonians describing the upper and lower layer for the edge states can be obtained as
\begin{equation}
  H_{u}' = \begin{pmatrix}
    k_y - i\kappa & \nu' \\
  \nu' & -k_y + i\kappa
  \end{pmatrix}.
\end{equation} 
\begin{equation}
  H_{l}' = \begin{pmatrix}
    -k_y + i\kappa & \nu' \\
  \nu' & k_y - i\kappa
  \end{pmatrix}.
\end{equation}
$\kappa$ is the skin depth of the edge states. Using the same method as in the previous subsection, i.e., the dynamics of an initial wavepacket $\Phi(t=0)=\left(\Phi^\uparrow,\Phi^\downarrow\right)^T$ in one period $T$ can be expressed as $U'(t) \Phi(t=0)$, where
\begin{equation}
  U'(t) = 
  \begin{cases}
    \exp(-iH_{u}'t), & 0<t<T/2\\
    \exp(-iH_{l}'(t-T/2))\exp(-iH_{u}'T/2), & T/2<t<T.
  \end{cases}\label{eq:Ueff2}
\end{equation}
In this way, we can get the wavepacket after one period $T$ as $\Phi(t=T) = U'(t=T) \Phi(t=0)$. If we denote 
\begin{equation}
  \begin{aligned}
  \boxed{\omega'^2 = \nu'^2+k_y^2 -\kappa^2-2ik_y\kappa}
  \end{aligned}
\end{equation}
and $A=-1+\exp(i\omega' T)$, the final state $\Phi'(t=T)$ can be expressed as
\begin{equation}
  \begin{aligned}
  \phi_1'(T)&= \Phi^\uparrow\left[\frac{e^{-i \omega' T} \left( A^2 \nu'^2 + 2 e^{i \omega' T} \omega'^2 \right)}{2 \omega'^2}\right]+\Phi^\downarrow \left[-\frac{A e^{-i \omega' T} \nu' \left( A k_y - i A \kappa + (A+2) \omega' \right)}{2 \omega'^2}\right], \\
  \phi_2'(T)&=\Phi^\uparrow\left[-\frac{A e^{-i \omega' T} \nu' \left( -A k_y + i A \kappa + (A+2) \omega' \right)}{2 \omega'^2}\right]+\Phi^\downarrow\left[\frac{e^{-i \omega' T} \left( A^2 \nu'^2 + 2 e^{i \omega' T} \omega'^2 \right)}{2 \omega'^2}\right]
  \end{aligned}
\end{equation}
Then, let's denote  the effective Hamiltonian for the edge states can be obtained as
\begin{equation}
  H_{\text{eff}}'=\begin{pmatrix}
    \frac{e^{-i \omega' T} \left( A^2 \nu'^2 + 2 e^{i \omega' T} \omega'^2 \right)}{2 \omega'^2}  & -\frac{A e^{-i \omega' T} \nu' \left( A k_y - i A \kappa + (A+2) \omega' \right)}{2 \omega'^2} \\
    -\frac{A e^{-i \omega' T} \nu' \left( -A k + i A \kappa + (A+2) \omega' \right)}{2 \omega'^2} & \frac{e^{-i \omega' T} \left( A^2 \nu'^2 + 2 e^{i \omega' T} \omega'^2 \right)}{2 \omega'^2}
  \end{pmatrix}
\end{equation}
Then eigenvalues are
\begin{equation}
  \begin{aligned}
E_{\text{eff}}'&=\frac{e^{-i \omega' T}}{2 \omega'^2}\left[A^2 \nu'^2 + 
2 e^{i \omega' T} \omega'^2 \pm
A\nu'\sqrt{-A^2 k_y^2  + 2 i A^2 k_y \kappa  + A^2 \kappa^2  +  (A+2)^2 \omega'^2}\right]\\
% &=\frac{e^{-i \omega' T}}{2 \omega'^2}\left[ 
%   A^2 \nu'^2 + 
% 2 e^{i \omega' T} \omega'^2 \pm
% A\nu'\sqrt{-A^2 k_y^2  + 2 i A^2 k \kappa  + A^2 \kappa^2  +  (A+2)^2 \left(\nu'^2+k_y^2-\kappa^2-2ik_y\kappa\right)}
% \right]\\
% &=\frac{e^{-i \omega' T}}{2 \omega'^2}\left[ 
%   A^2 \nu'^2 + 
% 2 e^{i \omega' T} \omega'^2 \pm
% 2A\nu'\omega'\sqrt{(A+1)}
% \right]\\
% &=\frac{e^{-i \omega' T}}{2 \omega'^2}\left[ 
%   A^2 \nu'^2 + 
% 2 e^{i \omega' T} \omega'^2 \pm
% 2A\nu'\omega'e^{i \omega' T/2}
% \right]\\
% &=\frac{A^2\nu'^2e^{-i \omega' T}}{2 \omega'^2}+1 \pm \frac{A\nu'}{\omega'}e^{-i \omega' T/2}\\
&=\left[1\pm\frac{A\nu'}{2\omega'}e^{-i \omega' T/2}\right]^2+\left(\frac{A\nu'}{2\omega'}e^{-i \omega' T/2}\right)^2.
  \end{aligned}
\end{equation}

Let's denote 
\begin{equation}
\boxed{\Omega=\frac{A\nu'}{2\omega'}e^{-i \omega' T/2}=\frac{\nu' }{2\omega'}\left[\exp(i\omega' T/2)-\exp(-i\omega' T/2)\right]= \frac{i\nu' \sin(\omega' T/2)}{\omega'}}
\end{equation}
Then the eigenenergies are
\begin{equation}
  E_{\text{eff}}' =(1\pm\Omega)^2+\Omega^2= 1 \pm 2 \Re\Omega + 2 (\Re\Omega)^2  - 2 (\Im\Omega)^2+i\left[  4  \Re\Omega \Im\Omega\pm 2  \Im\Omega\right]
\end{equation}

To measure the imaginary part of the edge spectrum, we can then retrieve Im$E_{\text{eff}}'$ as
\begin{equation}
  \boxed{\Im E_{\text{eff}}' =2\Im\Omega(2\Re\Omega\pm 1)}\label{eq:ImEeff2}
\end{equation}
This $\Im E_{\text{eff}}'$ can be exactly obtained by calculating $\Im\Omega$ and $\Re\Omega$,
\begin{equation}
  \boxed{
  \begin{aligned}
    \Im\Omega &= \frac{\left(\Re\omega'\right) \nu' \cosh\left(\frac{\left(\Im\omega'\right) T}{2}\right) \sin\left(\frac{\left(\Re\omega'\right) T}{2}\right)}{\abs{\nu'^2+k_y^2-\kappa^2-2ik_y\kappa}} + 
    \frac{\left(\Im\omega'\right) \nu' \cos\left(\frac{\left(\Re\omega'\right) T}{2}\right) \sinh\left(\frac{\left(\Im\omega'\right) T}{2}\right)}{\abs{\nu'^2+k_y^2-\kappa^2-2ik_y\kappa}}\\
    \Re\Omega &= \frac{\left(\Im\omega'\right) \nu' \cosh\left(\frac{\left(\Im\omega'\right) T}{2}\right) \sin\left(\frac{\left(\Re\omega'\right) T}{2}\right)}{\abs{\nu'^2+k_y^2-\kappa^2-2ik_y\kappa}} - 
    \frac{\left(\Re\omega'\right) \nu' \cos\left(\frac{\left(\Re\omega'\right) T}{2}\right) \sinh\left(\frac{\left(\Im\omega'\right) T}{2}\right)}{\abs{\nu'^2+k_y^2-\kappa^2-2ik_y\kappa}}
  \end{aligned}}\label{eq:ImReOmega}
\end{equation}

\subsubsection{Example}

\begin{figure}[H]
  \centering
  \includegraphics[width=0.9\textwidth]{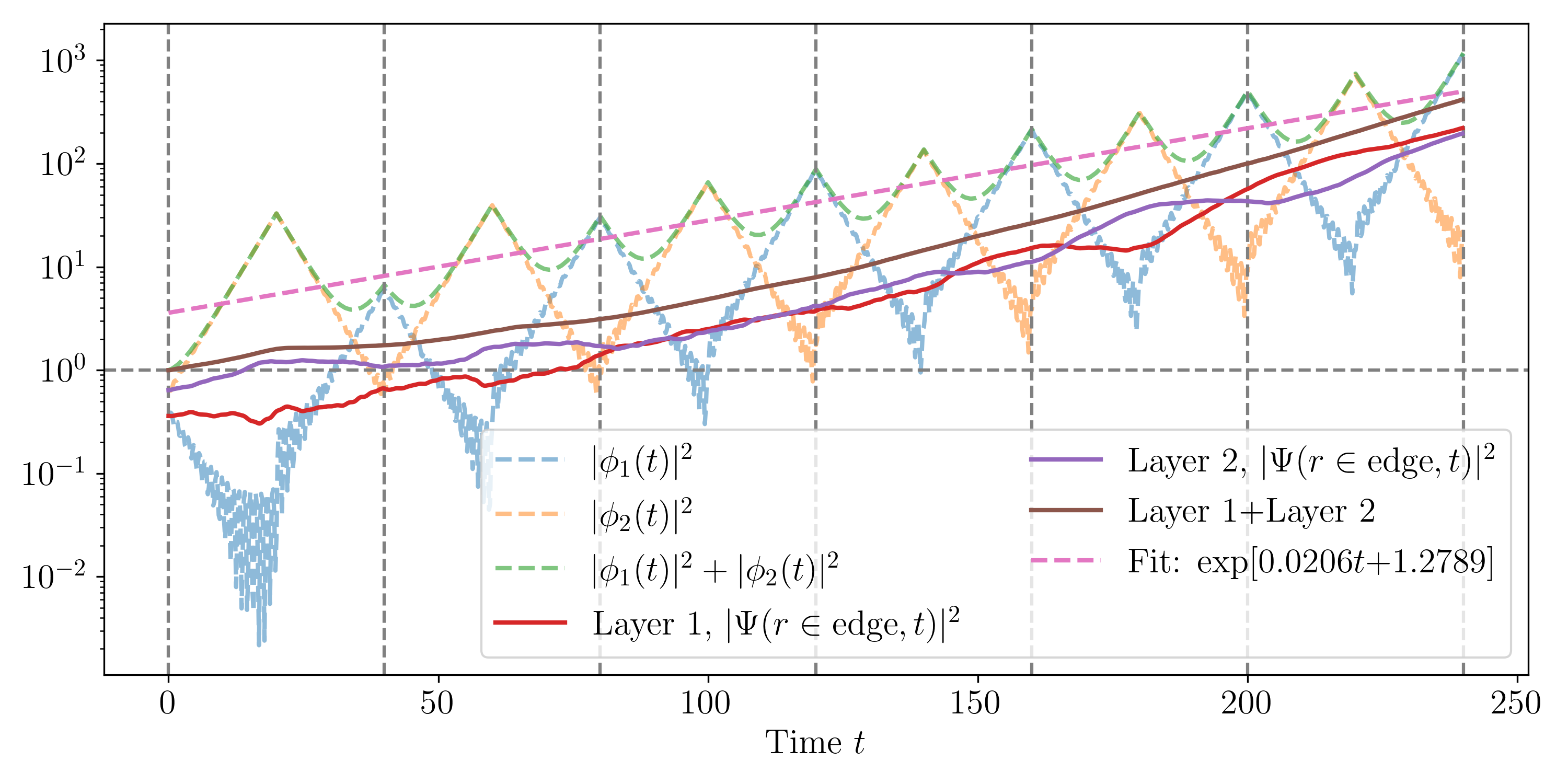}
  \caption{Time evolution of the wavepacket in the effective model for a single QSH layer with the opposite skin effect [Eq.~\eqref{eq:Ueff2}]. The initial state is $\Phi(t=0) = (0.6, 0.8)^T$, with parameters $\nu'=0.2$, $k_yv=3$, and $\kappa v=0.1, v=0.96$. The wavepacket exhibits periodic motion with amplification over six periods, indicating exponential growth of the edge states. The red dashed curves show the exponential fit of the growth rate in the numerical dynamics. The estimated growth rate in the numerical results is about $0.41$, which is in good agreement with the analytical prediction of $0.34$, confirming the validity of the effective theory. Parameters in the actual system are $\gamma=0.02$, $\nu=0.2$, $L_x=14$, and $L_y=4$.}
  \label{fig:eff2}
\end{figure}

To verify the analytical results in $\Im E_{\text{eff}}'$, we conduct numerical simulations in the effective model in a single QSH layer with the opposite skin effect. The initial state is set as $\Phi(t=0) = (0.6, 0.8)^T$, with parameters $\nu'=0.2$, $k_y v=3$, and $\kappa=0.1$. The time evolution $\Phi(t)=(\phi_1(t),\phi_2(t))^T=U'(t)\Phi(t=0)$ [Eq.~\eqref{eq:Ueff2}] of the wavepacket over six periods is shown in Fig.~\ref{fig:eff2}. It is observed that the wavepacket undergoes periodic motion with amplification. This confirms that in a single QSH layer with the opposite skin effect, the edge states exhibit exponential growth.

We further exponentially fit the growth rate in the dynamics of the numerically calculated effective model, and the results are shown in Fig.~\ref{fig:eff2} in red dashed curves. The growth rate in one period in the numerical results is about $0.0206/2\times T=0.41$. In the analytical results, we can estimate the growth rate in Eq.~\eqref{eq:ImEeff2} as $\Im E_{\text{eff}}' = 2\Im\Omega(2\Re\Omega\pm 1)$. With the parameters $\nu'=0.2$, $k_y v=3$, and $\kappa=0.1$, we can calculate $\Im\Omega\approx-0.1138$ and $\Re\Omega\approx-0.2132$ in Eq.~\eqref{eq:ImReOmega} with $\Im\omega'=\Im\sqrt{\nu'^2+k_y^2-\kappa^2-2ik_y\kappa}\approx -0.0998$ and $\Re\omega'=\Re\sqrt{\nu'^2+k_y^2-\kappa^2-2ik_y\kappa}\approx 3.0067$. The growth rate in the analytical results is $\Im E_{\text{eff}}' =2\Im\Omega(2\Re\Omega\pm 1)=2\times(-0.1138)\times(2\times -0.2132\pm 1)=-0.13$ and 0.34. In the long time limit, the growth rate in the numerical results is about 0.41, which is close to the analytical results of 0.34. This confirms the validity of our effective theory.

\begin{figure}
  \centering
  \includegraphics[width=0.8\textwidth]{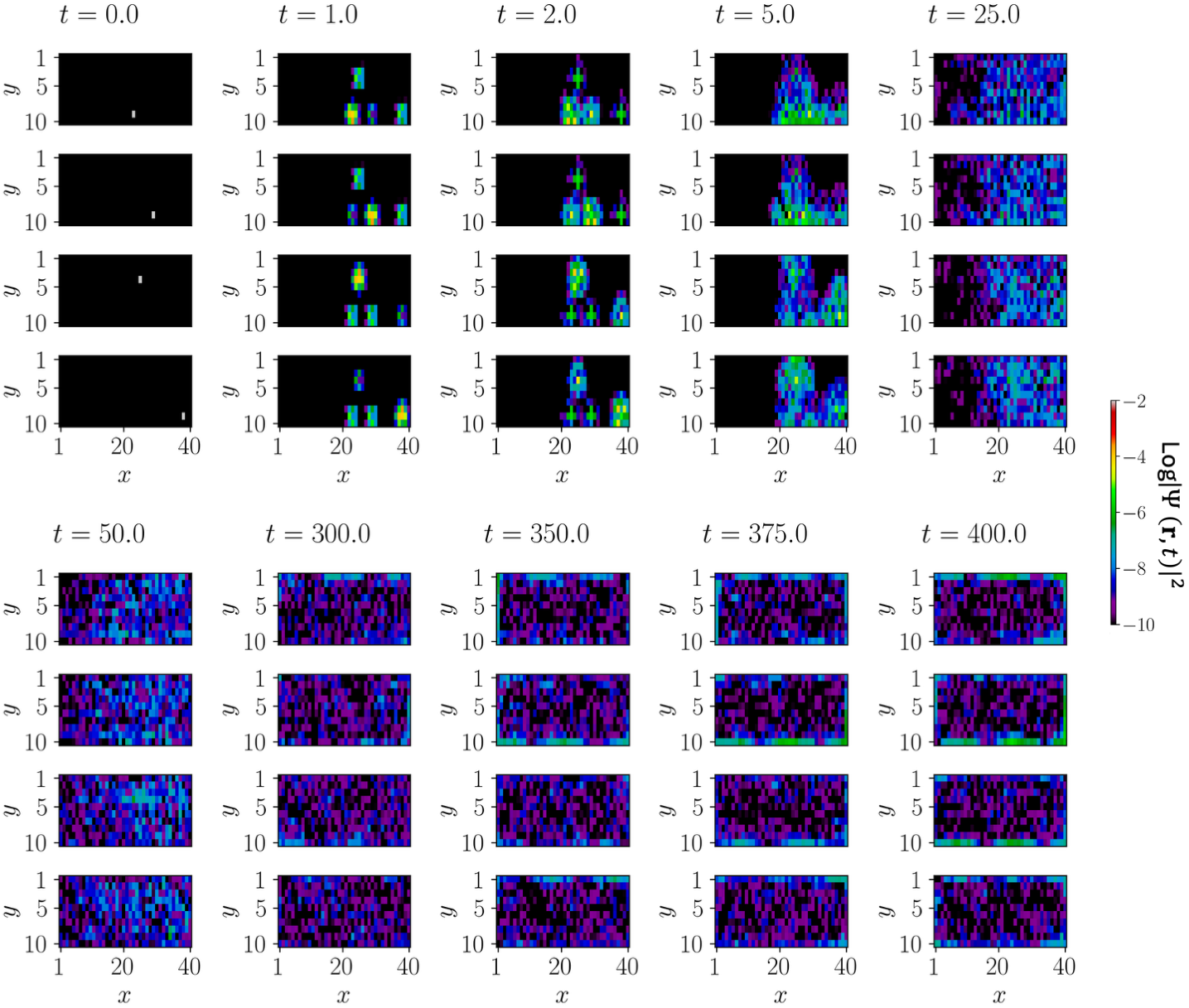}
  \caption{Non-directional spreading of the wavepacket in the stacked-QSH model at early times ($t<400$). Initially prepared as a bulk state, the wavepacket disperses isotropically and non-directionally throughout the bulk, prior to feedback-driven directional propagation and eventual localization at the edge at later times. Parameters used are the same as those in Fig.~2 of the main text: $\gamma=0.02$, $L_x\times L_y=40\times 10$, and $\mu=\nu=0.5$. }
  \label{fig:non_directional_spreading}
\end{figure}

\section{Bulk and edge point-gap winding and feedback-enforced edge amplification}

{In this section we detail how we compute the non-Hermitian point-gap winding numbers for the stacked quantum spin Hall (stacked-QSH) model introduced in Eqs.~(1) and (3) of the main text, and how these winding numbers diagnose the selective amplification of edge modes. The labels ``bulk'' and ``edge'' refer to different spectral constructions: the former uses the periodic Bloch Hamiltonian, whereas the latter uses the surface Green's function of a semi-infinite geometry; they are not obtained by partitioning the eigenvalues of one finite open system.}

\subsection{Point-gap winding in non-Hermitian band theory}

For a non-Hermitian Bloch Hamiltonian $H(\mathbf k)$, the relevant topological invariant associated with a \emph{point gap} at a base energy $E_\ast \in \mathbb C$ is the \emph{point-gap winding}
\begin{equation}
w(E_\ast)
=
\frac{1}{2\pi i}
\oint_{\mathcal C}
\partial_k
\log\det\bigl[H(k)-E_\ast\bigr]\, dk,
\label{eq:supp_w_def}
\end{equation}
where $\mathcal C$ is a closed loop in the Brillouin zone (BZ), $k$ is the coordinate along $\mathcal C$, and the logarithm is defined with a continuous branch along the loop. A nonzero integer $w(E_\ast)\in\mathbb Z$ signals that the spectrum winds nontrivially around $E_\ast$ as $k$ traverses the BZ, i.e.\ $E_n(k)$ encircles $E_\ast$ in the complex plane.

In our stacked-QSH model, the non-Hermitian drive is moderate and the spectrum exhibits a well-defined point gap around $E_\ast\simeq 0$ under both periodic and open boundary conditions. We therefore choose $E_\ast=0$ as the reference energy in the following.

\subsection{Bulk point-gap winding under periodic boundary conditions}

To characterize the bulk, we impose periodic boundary conditions (PBC) in both directions and consider the Bloch Hamiltonian
\begin{equation}
H_{\text{stacked-QSH}}(k_x,k_y)
\label{eq:supp_H_bulk}
\end{equation}
from Eq.~(2) of the main text. We evaluate the winding along a one-dimensional loop at fixed $k_y$; throughout this work we use $k_y=0$, but any loop $\mathcal C$ that is homotopically equivalent within the point-gap phase yields the same integer.

We discretize the loop $k_x\in[-\pi,\pi]$ as $k_x\to k_j$ ($j=1,\dots,N_k$) with $k_{N_k+1}\equiv k_1+2\pi$, and define
\begin{equation}
z_j=\det\bigl[H_{\text{stacked-QSH}}(k_j,0)-E_\ast\bigr].
\label{eq:supp_zj_def}
\end{equation}
To avoid numerical instabilities when $E_\ast$ approaches the spectrum, we introduce a small imaginary regulator and in practice compute
\begin{equation}
z_j=\det\bigl[H_{\text{stacked-QSH}}(k_j,0)-(E_\ast+i\eta)\bigr],
\qquad
\eta\sim 10^{-6}\text{--}10^{-4}.
\label{eq:supp_zj_reg}
\end{equation}
The bulk point-gap winding is then obtained from the discretized phase winding of $z_j$ as
\begin{equation}
\omega_{\rm bulk}(E_\ast)
=
\frac{1}{2\pi}
\sum_{j=1}^{N_k}
\mathrm{unwrap}\!\left[
\arg z_{j+1}
-
\arg z_j
\right]
\in\mathbb Z,
\label{eq:supp_w_bulk_disc}
\end{equation}
where $\mathrm{unwrap}$ denotes standard phase unwrapping to remove artificial $2\pi$ jumps arising from the principal branch of the argument. In all our numerics, the result is converged for $N_k\gtrsim 200$.

We choose $E_\ast$ inside the largest point gap by maximizing the minimal singular value
\begin{equation}
\Delta_{\rm bulk}(E)
=
\min_{k_x,k_y}\sigma_{\min}\bigl[H_{\text{stacked-QSH}}(k_x,k_y)-E\bigr],
\label{eq:supp_gap_bulk}
\end{equation}
over a small set of candidate energies; in the parameter regime relevant for the main text, $E_\ast=0$ is consistently found to lie deep inside the bulk point gap.

\subsection{Edge point-gap winding from the surface Green’s function}

To probe the boundary topology, we open the system along $y$ while keeping $k_x$ periodic. The stacked-QSH Hamiltonian can then be written in a principal-layer form
\begin{equation}
H(k_x)
=
\sum_{y}
\left[
c_y^\dagger H_0(k_x)\, c_y
+
c_{y+1}^\dagger T(k_x)\, c_y
+
c_y^\dagger T^\dagger(k_x)\, c_{y+1}
\right],
\label{eq:supp_pl}
\end{equation}
where $c_y$ annihilates the $8$-component spinor on layer $y$ (two Chern sectors, two spins, and two QSH copies), $H_0(k_x)$ is the onsite block obtained by setting $k_y=0$ in $H_{\text{stacked-QSH}}(k_x,k_y)$ (including inter-/intra-layer couplings and non-Hermitian drives), and $T(k_x)$ is the forward hopping block arising from the decomposition of the $k_y$ dependence,
\begin{equation}
\cos k_y\,\sigma_x+\sin k_y\,\sigma_z
=
e^{ik_y}T + e^{-ik_y}T^\dagger,
\label{eq:supp_T_def}
\end{equation}
lifted to the full $8\times 8$ space.

We consider a semi-infinite geometry $y\ge 0$ with a top edge at $y=0$. The corresponding surface Green’s function $g_s(k_x,E)$ satisfies
\begin{equation}
g_s(k_x,E)
=
\left[
(E+i\eta)I
-
H_0(k_x)
-
\Sigma(k_x,E)
\right]^{-1},
\qquad
\Sigma(k_x,E)
=
T(k_x)\,g_s(k_x,E)\,T^\dagger(k_x),
\label{eq:supp_gs_def}
\end{equation}
where $\Sigma$ is the self-energy due to the semi-infinite bulk and $\eta>0$ is a small positive regulator. We compute $g_s$ using the standard Sancho–Rubio iterative scheme:
\begin{equation}
\begin{aligned}
&g^{(0)}(k_x,E)
=
\bigl[(E+i\eta)I-H_0(k_x)\bigr]^{-1},
\quad
a^{(0)}(k_x)=T(k_x),
\quad
b^{(0)}(k_x)=T^\dagger(k_x),\\[0.3em]
&g^{(n+1)}(k_x,E)
=
\left[
(E+i\eta)I
-
H_0(k_x)
-
a^{(n)}(k_x)\,g^{(n)}(k_x,E)\,b^{(n)}(k_x)
\right]^{-1},\\[0.3em]
&a^{(n+1)}(k_x)
=
a^{(n)}(k_x)\,g^{(n)}(k_x,E)\,a^{(n)}(k_x),
\qquad
b^{(n+1)}(k_x)
=
b^{(n)}(k_x)\,g^{(n)}(k_x,E)\,b^{(n)}(k_x),
\end{aligned}
\label{eq:supp_SR_iter}
\end{equation}
until convergence $\|g^{(n+1)}-g^{(n)}\|_{\rm F}<{\tt tol}$ (we use ${\tt tol}\sim 10^{-12}$). The converged surface Green’s function is $g_s=g^{(n_\ast)}$.

The \emph{edge} point-gap winding is then defined as the phase winding of $\det[g_s^{-1}(k_x,E_\ast)]$ along the BZ:
\begin{equation}
\omega_{\rm edge}(E_\ast)
=
\frac{1}{2\pi}
\sum_{j=1}^{N_k}
\mathrm{unwrap}\Bigl[
\arg\det g_s^{-1}(k_{x,j+1},E_\ast)
-
\arg\det g_s^{-1}(k_{x,j},E_\ast)
\Bigr]
\in\mathbb Z,
\label{eq:supp_w_edge_def}
\end{equation}
where, as before, we discretize $k_x\in[-\pi,\pi]$ as $k_{x,j}$ and apply phase unwrapping to remove spurious $2\pi$ jumps. Since $\det g_s^{-1}=(\det g_s)^{-1}$, one may equivalently use $-\arg\det g_s$ in Eq.~\eqref{eq:supp_w_edge_def}.

\subsection{Representative values for the stacked-QSH model}

For the representative parameters used in the main text (e.g.\ those of Fig.~2), and with $E_\ast=0$ and $N_k=256$, we obtain
\begin{equation}
\boxed{
\omega_{\rm bulk}(E_\ast=0)=0,
\qquad
\omega_{\rm edge}(E_\ast=0)=2.
}
\label{eq:supp_w_results}
\end{equation}
The vanishing bulk winding $\omega_{\rm bulk}=0$ indicates that the periodic system lies in a \emph{topologically trivial point-gap phase}: the bulk bands do not wind around $E_\ast$ in the complex plane, consistent with the nearly real bulk spectrum observed in Fig.~1(c3) of the main text.

In contrast, the nonzero edge winding $\omega_{\rm edge}=2$ reveals that the boundary transfer problem has a nontrivial point gap of degree two. This implies the existence of a generalized Brillouin zone with $|\beta|\neq 1$ (a non-Hermitian skin effect) and, consequently, an inherent bias of the imaginary parts of the edge eigenvalues: two edge channels are systematically more amplifying ($\operatorname{Im}E>0$) than the bulk modes.

\color{black}

\section{Non-directional spreading of the wavepacket in the stacked-QSH model initially within time $t<400$}

In Fig.2 of the main text, we demonstrated the feedback-enforced chiral edge dynamics in our non-Hermitian stacked-QSH system [Eq.(3) in the main text] for parameters $\gamma=0.02$, lattice dimensions $L_x	\times L_y=40	\times 10$, and coupling strengths $\mu=\nu=0.5$ (intermediate coupling regime). These couplings respectively connect layers exhibiting parallel and anti-parallel non-Hermitian skin effects (NHSE) [see Fig.~1(c1)]. Specifically, in Fig.~2(a1), we showed the propagation of a wavepacket initially prepared in a bulk state, evolving according to $\Psi(\bold r, t)=\exp(-iH_{\text{stacked-QSH}}t)\Psi(\bold r\notin\text{Edge}, t=0)$, which eventually localizes at the system edge at long times. Notably, during the initial transient period ($t<400$), the wavepacket exhibits non-directional spreading within the bulk region, before subsequently becoming localized at the edge.

\section{Reality of edge spectra under $x$- and $y$-OBC in NH Chern and NH QSH insulators}

In the main text, we emphasized that, on a rectangular sample with open boundaries in both $x$ and $y$, the topological edge modes of the non-Hermitian Chern and QSH models remain (essentially) real, even though the bulk spectrum is genuinely non-Hermitian [see Fig.~1(a3,b3) and the discussion around Eq.~(1)]. This property underlies our statement that the edge channels can be viewed as non-Hermitian deformations of their Hermitian counterparts, whereas bulk modes acquire complex energies. For completeness, we now give a self-contained real-space argument showing that the relevant OBC Hamiltonians are pseudo-Hermitian and possess an unbroken global $\mathcal{PT}$ symmetry that pins all edge eigenvalues to the real axis.

In the real-space representation, the Hamiltonian on a rectangular sample with OBC in both $x$ and $y$ takes the schematic form
\begin{equation}
H = \left(m + \cos k_x + \cos k_y\right) \sigma_x + \left(i\gamma_j + \sin k_x\right) \sigma_y + \sin k_y \sigma_z
= H_{\rm herm} + i\gamma(\sigma_y\otimes \mathbb{I}),
\end{equation}

\paragraph*{NH Chern insulator.}
\textbf{(i) Pseudo-Hermiticity and a global $\mathcal{PT}$ symmetry.}
Let $\mathcal{P}$ denote inversion about the sample center, mapping a lattice site $(i,j)$ to $(L_x-1-i,L_y-1-j)$. Define the antiunitary operator
\begin{equation}
\Theta \equiv \mathcal{PT} = (\sigma_x \otimes \mathcal{P})K,
\end{equation}
with $K$ complex conjugation. Using $\sigma_x\sigma_y\sigma_x=-\sigma_y$, the reality of the hopping blocks in $H_{\rm herm}$, and the fact that the rectangle is inversion symmetric, one directly verifies the two symmetry relations
\begin{equation}
(\sigma_x\otimes\mathbb{I})H(\sigma_x\otimes\mathbb{I}) = H^\dagger,
\qquad
\Theta H\Theta^{-1}=H.
\end{equation}
Thus $H$ is pseudo-Hermitian and commutes with $\Theta$. Standard results of $\mathcal{PT}$-symmetric/pseudo-Hermitian quantum mechanics imply: eigenvalues are either real or occur in complex-conjugate pairs. Moreover, any eigenstate that is itself an eigenstate of $\Theta$ must have a strictly real eigenvalue.

\textbf{(ii) Edge states under $x$-OBC and $y$-OBC are $\mathcal{PT}$ eigenstates.}
Open boundaries in both directions produce pairs of counter-localized edge modes supported on opposite, inversion-related edges. Because the finite rectangle is globally inversion symmetric, $\mathcal{P}$ exchanges the two edges. As a result, the Hamiltonian mixes them into symmetric/antisymmetric combinations that can be chosen to be simultaneous eigenstates of $\Theta$:
\begin{equation}
\Theta \ket{\psi_{\rm edge}} = e^{i\phi}\ket{\psi_{\rm edge}} .
\end{equation}
For such (unbroken) $\mathcal{PT}$-symmetric states one has
\begin{equation}
H\Theta\ket{\psi_{\rm edge}}=\Theta H\ket{\psi_{\rm edge}}
=\Theta\big(E\ket{\psi_{\rm edge}}\big)
=E^* \Theta\ket{\psi_{\rm edge}},
\end{equation}
and since $\Theta\ket{\psi_{\rm edge}} \propto \ket{\psi_{\rm edge}}$, it follows that $E=E^*$, i.e.\ the edge eigenvalue is strictly real. This matches our OBC$\times$OBC numerics, where edge levels remain real across the parameter range where the edge states exist.

\paragraph*{NH QSH insulator.}
In our NH QSH implementation (two time-reversal-related copies of the Chern model), with a uniform non-Hermitian term $i\gamma\sigma_y\otimes s_0$ and a rectangular sample with open boundaries in both $x$ and $y$ that is inversion-symmetric about its center, the edge energies are strictly real.

Write the real-space Hamiltonian as
\begin{equation}
H=H_{\rm herm}+i\gamma(\sigma_y\otimes s_0\otimes \mathbb{I}_{\rm lat}),
\end{equation}
where $H_{\rm herm}$ is composed of real hopping blocks plus their Hermitian conjugates. Define the antiunitary operator
\begin{equation}
\Theta=(\sigma_x\otimes s_0\otimes \mathcal{P})K,
\end{equation}
with $\mathcal{P}$ the inversion about the sample center and $K$ complex conjugation. Using $\sigma_x\sigma_y\sigma_x=-\sigma_y$ and the reality of $H_{\rm herm}$, one verifies
\begin{equation}
(\sigma_x\otimes s_0)H(\sigma_x\otimes s_0)=H^\dagger,\qquad \Theta H\Theta^{-1}=H,
\end{equation}
so $H$ is pseudo-Hermitian and $\Theta$-symmetric. Consequently, eigenvalues are either real or appear in complex-conjugate pairs; any eigenstate that is itself a $\Theta$ eigenstate has a strictly real eigenvalue.

Under $x$-OBC and $y$-OBC, inversion $\mathcal{P}$ exchanges the two opposite edges, so the Hamiltonian mixes them into symmetric/antisymmetric combinations that can be chosen as simultaneous eigenstates of $\Theta$. Hence every edge state satisfies $\Theta\ket{\psi_{\rm edge}}=e^{i\phi}\ket{\psi_{\rm edge}}$ and its energy is real. Equivalently, since $\Theta(\sigma_y\otimes s_0)\Theta^{-1}=-(\sigma_y\otimes s_0)$, we have $\langle\sigma_y\otimes s_0\rangle=0$ on each edge state and the non-Hermitian onsite term contributes no imaginary shift.

The time-reversal symmetry $\mathcal{T}=(\mathbb{I}_\sigma\otimes i s_y\otimes\mathbb{I})K$ with $\mathcal{T}^2=-1$ remains intact (the non-Hermitian term is $\mathcal{T}$-even), producing Kramers pairs but not affecting the reality of edge energies.

\section{Possible experimental realization in a classical topoelectrical circuit}

In our main text, we proposed a feedback-enforced stacked-QSH architecture and showed that its key phenomenology can be distilled into three experimentally accessible ingredients: (i) non-Hermitian pumping; (ii) a four-layer structure with interlayer tunneling and oppositely directed pumping; and (iii) chiral topological pumping. Importantly, each of these building blocks has already been demonstrated in established platforms. Specifically, (i) has been realized in electrical circuits~\cite{helbig2020generalized,liu2021non,zou2021observation}, {photonic quantum walks~\cite{weidemann2020topological,xiao2020non}}, mechanical metamaterials~\cite{ghatak2020observation}, ultracold atoms~\cite{li2020topological,liang2022dynamic} and quantum circuits~\cite{smith2019simulating,gou2020tunable,koh2022simulation,kirmani2022probing,frey2022realization,chertkov2023characterizing,chen2023robust,yang2023simulating,iqbal2023creation,shen2023observation,koh2023observation}. Ingredient (ii) can typically be realized whenever the platform supports coupled bilayers, which is feasible in electrical circuits~\cite{hofmann2019chiral,ezawa2019electric,helbig2020generalized,hofmann2020reciprocal,liu2020gain,liu2021non,stegmaier2021topological,zhang2020non,zhang2022observation,shang2022experimental,yuan2023non,zou2023experimental,zhu2023higher,zhang2023electrical}, suitable photonic crystals~\cite{pan2018photonic,xiao2020non,zhu2020photonic,song2020two,ao2020topological} and quantum circuits~\cite{smith2019simulating,gou2020tunable,koh2022simulation,kirmani2022probing,frey2022realization,chertkov2023characterizing,chen2023robust,yang2023simulating,iqbal2023creation,shen2023observation,koh2023observation}. Finally, (iii) has also been realized in various photonic~\cite{weidemann2020topological}, mechanical~\cite{ghatak2020observation} and acoustic systems~\cite{zhang2021acoustic}, as well as in cold atoms~\cite{liang2022dynamic} and quantum circuits~\cite{roushan2014observation}. {As such, suitably designed photonic systems, electrical circuits, and quantum circuits on digital quantum processors can realize our model and its feedback-enforced bulk suppression and edge amplification using existing experimental technology.}

While appropriately designed metamaterial platforms are in principle all feasible, we recommend using electrical circuits due to their versatility~\cite{lee2018topolectrical,hofmann2019chiral}. In general, it is feasible to construct a tailored topolectrical circuit to simulate a non-Hermitian Chern insulator for each layer, and subsequently couple them through mutual inductances. {By doing so, the bulk-real/edge-amplified regime can be detected by comparing the impedance response of bulk and boundary nodes across the circuit.} Below, we provide a brief proposal of how our model can be realized and observed in such a system. 

Below, we first present the INIC-based realization of a \emph{single Chern layer exhibiting the non-Hermitian skin effect}, and then show how to couple four such layers via site-preserving ($\mu$) and sublattice-inverting ($\nu$) capacitive links to realize the stacked-QSH architecture with feedback-enforced helical edge transport.

\subsection{Single-layer realization of a Chern insulator with NHSE using INICs}

\subsubsection{INIC building block and nonreciprocity}

\begin{figure}[h]
    \centering
    \includegraphics[width=0.7\textwidth]{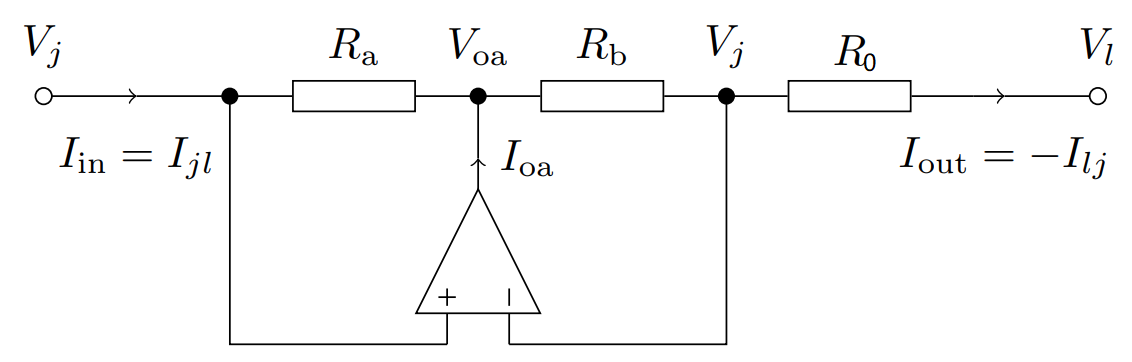}
    \caption{A sketch of an INIC circuit, adapted from Supplemental Material of Ref.~\cite{hofmann2019chiral}.}
    \label{rfig:INICb}
\end{figure}

The topological character and nonreciprocity in a single layer originate from the INIC next-nearest-neighbor same-sublattice couplings ($A$-$A$ and $B$-$B$)~\cite{hofmann2019chiral}. These elements break time-reversal symmetry and circuit reciprocity by effectively implementing negative and positive resistance in opposite directions. Let $\nu\equiv R_b/R_a$ denote the INIC resistance ratio, and let $R_0$ be the series resistor on the through path of the element. For the two-node link between sites $j$ and $l$ shown in Fig.~\ref{rfig:INICb}, assuming ideal op-amps (infinite input impedance, linear regime), the input and output currents satisfy
\begin{equation}
\begin{aligned}
I_{\text{in}}&=-\frac{R_b}{R_aR_0}(V_j-V_l),\\
I_{\text{out}}&=\frac{1}{R_0}(V_j-V_l).
\end{aligned}
\label{eq:INIC_currents}
\end{equation}
Writing this in Laplacian form yields the asymmetric two-port admittance
\begin{equation}
\begin{pmatrix}
I_{jl}\\[2pt] I_{lj}
\end{pmatrix}
=\frac{1}{R_0}
\begin{pmatrix}
-\nu & \nu\\[2pt]
-1 & 1
\end{pmatrix}
\begin{pmatrix}
V_j\\[2pt] V_l
\end{pmatrix},
\qquad \nu\equiv\frac{R_b}{R_a},
\label{eq:INIC_link_Lap}
\end{equation}
which is manifestly non-Hermitian and nonreciprocal.

\subsubsection{Lattice assembly and Pauli-form Laplacian}

\begin{figure}[h]
    \centering
    \includegraphics[width=0.96\textwidth]{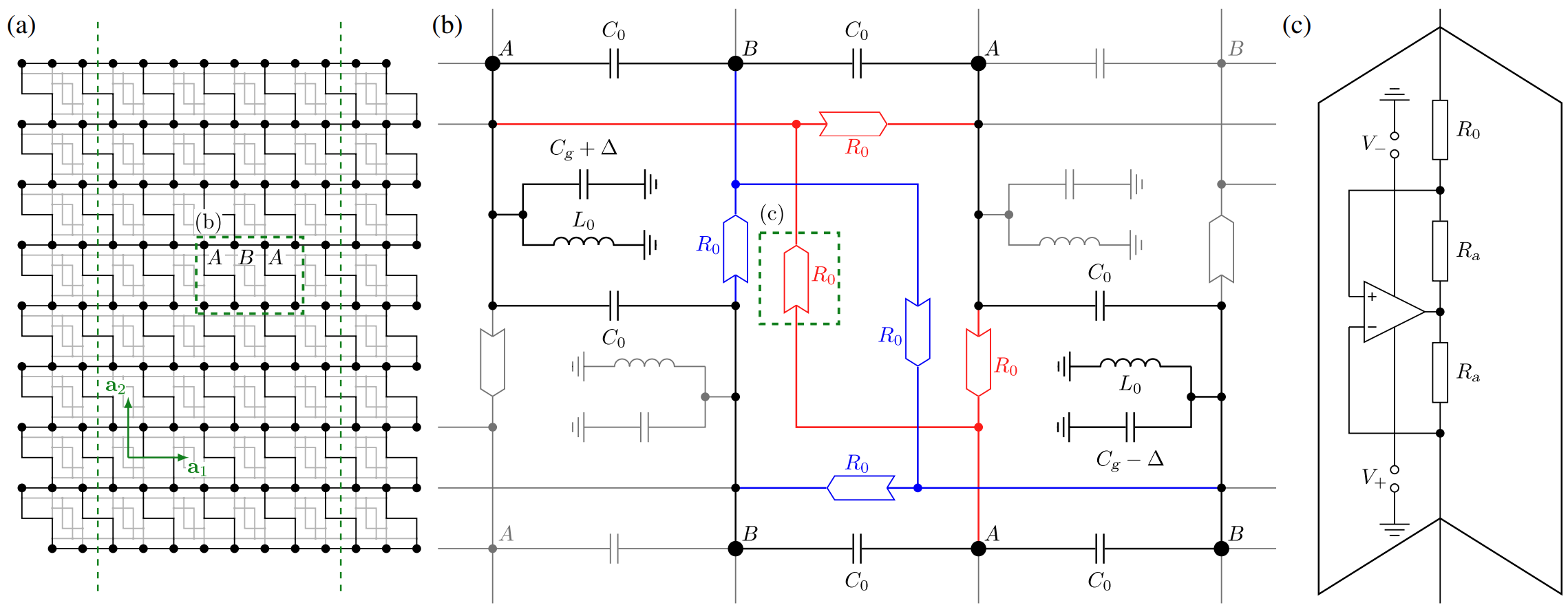}
    \caption{Topoelectrical Chern circuits, adapted from Ref.~\cite{hofmann2019chiral}. (a) A sketch of the topoelectrical Chern circuit. (b) The circuit unit cell consists of two nodes, each of which is connected to three adjacent nodes through a capacitor $C_0$ and to six next-nearest neighbors through INICs. (c) INICs.}
    \label{rfig:topoelec_chernb}
\end{figure}

We now assemble a two-node $(A,B)$ unit cell on a square lattice (brick-wall gauge). Nearest-neighbor AB capacitors $C_0$ implement the real hopping network; grounded capacitors $C_g\pm\Delta$ set on-site offsets; same-sublattice INICs implement the complex, direction-dependent next-nearest couplings. In Bloch form with $\psi=(V_A,V_B)^{\mathsf T}$ and $k=(k_x,k_y)$, the grounded circuit Laplacian reads
\begin{equation}
J_{\mathrm{TCC}}(k;\omega)=i\omega\left(J_0\mathbb{I}+J_x\sigma_x+J_y\sigma_y+J_z\sigma_z\right),
\label{eq:TCC_Pauli}
\end{equation}
with components (parameters labeled in Fig.~\ref{rfig:topoelec_chernb})
\begin{equation}
J_0=3C_0+C_g-\frac{1}{\omega^2 L_0}-i\frac{1}{\omega R_0}
\left(1-\frac{\nu_A+\nu_B}{2}\right)\Big[3-\cos k_x-\cos k_y-\cos(k_x-k_y)\Big],
\label{eq:J0}
\end{equation}
\begin{equation}
J_x=-C_0\Big[1+\cos k_x+\cos k_y\Big],
\label{eq:Jx}
\end{equation}
\begin{equation}
J_y=-C_0\Big[\sin k_x+\sin k_y\Big],
\label{eq:Jy}
\end{equation}
\begin{equation}
\begin{aligned}
J_z=\ &\Delta+\frac{1}{\omega R_0}\left(1+\frac{\nu_A+\nu_B}{2}\right)\Big[\sin k_x-\sin k_y-\sin(k_x-k_y)\Big]\\
&\ +i\frac{1}{\omega R_0}\frac{\nu_A-\nu_B}{2}\Big[3-\cos k_x-\cos k_y-\cos(k_x-k_y)\Big].
\end{aligned}
\label{eq:Jz}
\end{equation}
Here $\nu_A$ ($\nu_B$) is the INIC ratio on the $A$ ($B$) sublattice links. When $\nu_A=\nu_B=1$ the Laplacian is Hermitian (up to uniform loss), reproducing the baseline Chern circuit; small detuning $\nu_A\neq\nu_B$ produces a controlled non-Hermitian sector and an NHSE.

\subsubsection{Basis alignment to the target two-band form}

For comparison with the target two-band model, we employ the unitary
\begin{equation}
U=\frac{1}{\sqrt{2}}
\begin{pmatrix}
1 & -i\\[2pt]
i & -1
\end{pmatrix},
\qquad
U\sigma_xU^\dagger=-\sigma_x,\ \ U\sigma_yU^\dagger=\sigma_z,\ \ U\sigma_zU^\dagger=\sigma_y.
\label{eq:Urotate}
\end{equation}
At a chosen design frequency $\omega_0=2\pi f$, dividing the Laplacian by $i\omega_0 C_0$ defines a dimensionless form $\tilde J\equiv J/(i\omega_0 C_0)$, which we identify with the rotated single-layer target Hamiltonian
\begin{equation}
H'_{\mathrm{Ch}}=-(m+\cos k_x+\cos k_y)\sigma_x+(\sin k_y)\sigma_y+\big(i\gamma+\sin k_x\big)\sigma_z.
\label{eq:Hprime}
\end{equation}
This fixes the parameter map:
\begin{equation}
m\ \text{set by intra-/inter-cell AB weights and fine-tuned by }\Delta,\qquad
\gamma=\frac{\nu_A-\nu_B}{2R_0\omega_0C_0}.
\label{eq:map_m_gamma}
\end{equation}
The sign of $\gamma$ (and thus the NHSE direction) is controlled by $\nu_A\gtrless \nu_B$. The Chern number follows the standard two-band windows with band closings at $m=-2,0,2$; hence $C=\pm1$ for $m\in(-2,0)$ and $m\in(0,2)$ (sign depends on orientation), and $C=0$ otherwise.

\subsection{Coupling all layers:  $\mu,\nu$ implementation}

\subsubsection{Target architecture and layer assignment}

We realize four Chern layers, arranged as two QSH bilayers with \emph{opposite} NHSE directions. Concretely:
\begin{itemize}
\item Layer 1: choose $m$ in a $C=+1$ window and set $\nu_A>\nu_B$ so that $\gamma_1=+\gamma$.
\item Layer 2: choose $m$ in a $C=-1$ window and set $\nu_A>\nu_B$ so that $\gamma_1=+\gamma$.
\item Layer 3: choose the same $m$ window as Layer 1 ($C=+1$) but set $\nu_A<\nu_B$ so that $\gamma_2=-\gamma$.
\item Layer 4: choose the same $m$ window as Layer 2 ($C=-1$) and set $\nu_A<\nu_B$ so that $\gamma_2=-\gamma$.
\end{itemize}
This assignment yields two time-reversal-partner Chern copies per QSH bilayer and flips the NHSE sign between the two bilayers, which is essential for the feedback-enforced edge transport upon interlayer coupling.

\subsubsection{Interlayer couplings: intra-QSH $\mu$ and inter-QSH $\nu$}

All interlayer links are implemented as same-sublattice capacitors (A-A and B-B) at identical in-plane coordinates $(x,y)$. The intra-QSH coupling between Layers $1\leftrightarrow2$ and $3\leftrightarrow4$ is realized by capacitors $C_\mu$, and the inter-QSH coupling between Layers $1\leftrightarrow3$ and $2\leftrightarrow4$ is realized by capacitors $C_\nu$. At the design frequency,
\begin{equation}
\mu=\frac{C_\mu}{C_0},\qquad \nu=\frac{C_\nu}{C_0}.
\label{eq:mu_nu_ratios}
\end{equation}
In the global $(A,B)$ basis, the four-layer (dimensionless) Bloch Laplacian reads
\begin{equation}
\mathcal{J}(k)=
\begin{pmatrix}
\tilde J_{\mathrm{Ch}}(+C,+\gamma)& \mu\mathbb{I} & \nu\mathbb{I} & 0 \\
\mu\mathbb{I} & \tilde J_{\mathrm{Ch}}(-C,+\gamma)& 0 & \nu\mathbb{I} \\
\nu\mathbb{I} & 0 & \tilde J_{\mathrm{Ch}}(+C,-\gamma) & \mu\mathbb{I} \\
0 & \nu\mathbb{I} & \mu\mathbb{I} & \tilde J_{\mathrm{Ch}}(-C,-\gamma)
\end{pmatrix}.
\label{eq:block_four}
\end{equation}
Hardware-wise, both $\mu$ and $\nu$ are simply A-A/B-B capacitors; no sublattice inversion is required.

\subsection{Example parameters at $f=100~\mathrm{kHz}$ (explicit picks)}

Pick the following:
\begin{equation}
C_0=0.1~\mu\mathrm{F},\quad R_0=100~\Omega,\quad \omega_0=2\pi\times 100~\mathrm{kHz}.
\label{eq:base_numbers}
\end{equation}
Set INIC ratios (e.g., with $R_a=10~\mathrm{k}\Omega$):
\begin{equation}
\nu_A=1.125(R_{b,A}\approx 11.3~\mathrm{k}\Omega),\qquad
\nu_B=0.875(R_{b,B}\approx 8.66~\mathrm{k}\Omega),
\end{equation}
so that
\begin{equation}
\gamma=\frac{\nu_A-\nu_B}{2\omega_0 R_0}
=\frac{0.25/2}{\omega_0 R_0}\approx 2.0~\mathrm{nF}\approx 0.02C_0 .
\label{eq:gamma_pick}
\end{equation}
For interlayer couplings choice, in the dimensionless convention,
\begin{equation}
\mu=\nu=0.5\quad\Rightarrow\quad
C_\mu=C_\nu=0.5C_0=50~\mathrm{nF}.
\label{eq:munu_pick}
\end{equation}
Mass tuning: use a small grounded imbalance $\Delta=\pm 5~\mathrm{nF}$ to move $m$ across the critical values ($m=-2,0,2$) as needed.

\section{Possible quantum-circuit realization on a digital superconducting processor}

\subsection{Building blocks: chiral propagation + nonunitary pumping}

Our four-layer model can be compiled to a gate sequence on a universal superconducting quantum processor by combining two experimentally established primitives:
(i) \emph{interaction-induced chiral topological dynamics} that realizes effective Chern transport on a qubit chain via compressed 2D$\to$1D mapping with Trotter/VQA compilation~\cite{koh2022simulation}; and
(ii) \emph{programmable nonunitary evolutions} realized with ancilla-assisted post-selection (or engineered loss), which have been used to observe non-Hermitian transport and the skin effect on present-day devices~\cite{shen2025observation,shen2025observation,wang2024non}.
The former provides the unitary Chern layer per time step, while the latter implements our layer-resolved imaginary potential (“pumping”) that breaks reciprocity in a controlled way.

\paragraph{Encoded single-layer Chern step.}
Following Ref.~\cite{koh2022simulation}, we encode an effective two-band Chern lattice on a 1D register using the compressed two-particle mapping, and implement one Trotter micro-step as a brick-wall of single- and two-qubit blocks that realize the split Hamiltonian (single-particle hops and two-particle interactions).
For notational convenience we write one compiled Chern step as
\begin{equation}
U_{\mathrm{Ch}}(\delta t)\equiv e^{-i h_y\sigma_y \delta t}e^{-i h_x\sigma_x \delta t}e^{-i h_0 \mathbb{I} \delta t},
\end{equation}
with $h_{x,y}$ realized by nearest-neighbor entanglers and single-qubit phases as in~\cite{koh2022simulation}, and time-reversal breaking injected by phase-staggered two-qubit gates; this implements, up to a basis choice, the unitary part of Eq.~(\ref{eq:Hprime}) layer-by-layer.

\paragraph{Layer-resolved nonunitary pumping (axis-aligned to $ \sigma_y $).}
The non-Hermitian term in Eq.~(\ref{eq:Hprime}) is $i\gamma\sigma_y$.
We realize it by inserting after each $U_{\mathrm{Ch}}$ a local, sublattice-selective nonunitary pump
\begin{equation}
\tilde U_{\mathrm{pump}}(\gamma,\delta t)=R_x(-\tfrac{\pi}{2})\Big[\prod_{x}K_x(Z;\gamma,\delta t)\Big]R_x(+\tfrac{\pi}{2}),
\end{equation}
where $K_x(Z;\gamma,\delta t)\approx e^{+\gamma\delta t Z_x}$ is implemented via an ancilla-assisted Kraus map on site $x$:
\begin{equation}
K_{\pm}=\sqrt{1-p_\pm}|0\rangle\langle 0|+\sqrt{1-q_\pm}|1\rangle\langle 1| ,
\end{equation}
with $(p_\pm,q_\pm)$ chosen so that the effective map approximates $e^{+\gamma\delta tZ}$; the pre/post rotations $R_x(\mp\pi/2)$ rotate this into $e^{+\gamma\delta tY}$.
Operationally, we couple each system qubit to an ancilla through a controlled rotation and post-select on the ancilla outcome~\cite{shen2025observation}; as a hardware-native alternative, a dump resonator or Purcell-engineered lossy mode can enact amplitude damping at a programmable rate~\cite{wang2024non}.

\subsection{Register layout and four-layer stacking}

We realize four compiled Chern layers on four disjoint qubit registers $\mathcal{R}_{1\ldots 4}$ of equal length $N$.
Layers 1/2 form one QSH bilayer and Layers 3/4 form the other, with opposite NHSE directions: choose the Chern-mass windows so that $(C_1,C_2)=(+1,-1)$ and $(C_3,C_4)=(+1,-1)$, and set opposite pump signs by programming $\gamma_{1,2}=+\gamma$ and $\gamma_{3,4}=-\gamma$.
This realizes the “oppositely directed pumping” required for feedback-enforced helical transport.

\paragraph{Interlayer couplings \texorpdfstring{$\mu$}{mu} and \texorpdfstring{$\nu$}{nu}.}
In the effective block of Eq.~(\ref{eq:Hprime}), the intra-QSH coupling on each site enters as $V=\mu\mathbb{I}+\nu\sigma_x$ in sublattice space.
We therefore compile two ingredients between the corresponding qubits of different registers:
\begin{equation}
\boxed{U_\nu(\delta t)=\prod_{x=1}^{N}\Big[R_y^{(\ell)}(-\tfrac{\pi}{2})R_y^{(\ell'')}(-\tfrac{\pi}{2})e^{-i\nu\delta t Z_{x,\ell}Z_{x,\ell''}}R_y^{(\ell)}(\tfrac{\pi}{2})R_y^{(\ell'')}(\tfrac{\pi}{2})\Big]} \quad\Longleftrightarrow e^{-i\nu\delta t X\otimes X},
\end{equation}
realizing the $\sigma_x$ part, and a sublattice-diagonal piece that does not flip sublattices,
\begin{equation}
U_\mu(\delta t)=\prod_{x=1}^{N} e^{-i\mu\delta tZ_{x,\ell}Z_{x,\ell'}}\text{(or an equivalent diagonal synthesis)},
\end{equation}
which contributes as the $\mu\mathbb{I}$ block at the encoded level. Here $(\ell,\ell')=(1,2)$ or $(3,4)$ (intra-QSH) and $(\ell,\ell'')=(1,3)$ or $(2,4)$ (inter-QSH). In practice, cross-resonance/parametric iSWAP+phase blocks with single-qubit pre/post rotations synthesize these couplings within each Trotter step.

\subsubsection{One Trotter period and parameter map}

To reduce Trotter error we use a symmetric composition as
\begin{equation}
U_F \approx U_\mu\left(\tfrac{\delta t}{2}\right)U_\nu\left(\tfrac{\delta t}{2}\right)
\Big[\prod_{\ell=1}^4 \tilde U_{\mathrm{pump}}^{(\ell)}(\gamma_\ell,\delta t) U_{\mathrm{Ch}}^{(\ell)}(\delta t)\Big]
U_\nu\left(\tfrac{\delta t}{2}\right)U_\mu\left(\tfrac{\delta t}{2}\right),
\end{equation}
repeated $M$ times with $\delta t=T/M$.
Matching to Eq.~(\ref{eq:Hprime}) fixes
$m$ via the compiled imbalance in $h_x$ (fine-tuned by single-qubit $Z$ phases),
$\gamma$ via ancilla angles / damping rates and the $R_x$ basis rotations,
$\mu=\frac{\theta_\mu}{\delta t},\nu=\frac{\theta_\nu}{\delta t},$
where $\theta_{\mu,\nu}$ are the compiled two-qubit phases per step.

\subsubsection{Readout and signature of the PT threshold}

Initialize a wave packet on a boundary site of Layer~1, evolve under $U_F^M$, and measure site-resolved occupations on all four registers.
We diagnose the onset of feedback-stabilized helical edge transport and PT symmetry breaking by: 
(i) directed circulation of edge excitations across the four layers; 
(ii) growth and saturation of a layer-resolved edge-occupation imbalance; and 
(iii) the emergence of nonzero growth rates (local $\mathrm{Im}E$) and their redistribution between edge and bulk, consistent with the \emph{realness/threshold condition} governed by $\mu,\nu,\gamma$ (see Supplement for the analytical criterion).
These diagnostics parallel those in recent non-Hermitian chain experiments~\cite{shen2025observation}, but here are enforced and routed by the stacked-QSH feedback network.

\subsubsection{Feasibility and example scales}

State-of-the-art devices with mid-circuit measurement/reset and parallel two-qubit gates suffice.
As a concrete example, each layer with $N=10\text{-}16$ sites (total $40\text{-}64$ system qubits) plus one ancilla per $2\text{-}3$ sites (reused via reset) allows $M\sim50\text{-}100$ Trotter steps.
Chiral propagation follows the gate counts in~\cite{koh2022simulation}; nonunitary pumps use ancilla embeddings with one measurement per targeted site per few steps~\cite{shen2025observation}. 
To keep depth and sampling cost practical on NISQ hardware, we combine symmetric Trotter with variational recompilation and/or \emph{sparse pumping} (apply pumps every few steps). If available, hardware loss elements can replace post-selection to boost sampling efficiency~\cite{wang2024non}.

\section{Toward feedback-enforced edge-selected lasing}

The edge-biased pumping scheme of the synthetic magnetic-ring laser by Bandres~\cite{bandres2018topological} remains a milestone for single-mode topological lasing: uniformly distributed loss inside the lattice, combined with selective gain at the outer boundary, ensures that only the pre-existing edge channel acquires net amplification.  Our stacked non-Hermitian QSH model offers a different, potentially simpler route toward the same phenomenology in that uniform pumping already suffices to isolate a single edge super-mode.

As summarised in the main text, each QSH layer possesses balanced gain and loss $\pm i\gamma$; inter-layer couplings $\mu$ and $\nu$ hybridise the two copies.  A Schrieffer-Wolff reduction around the Bloch momentum most susceptible to complex eigenvalues yields an effective $2\times2$ PT dimer.  In this reduced description the entire bulk spectrum remains real whenever
\begin{equation}
  \mu\nu \ge \gamma.
  \label{eq:reality_condition}
\end{equation}
Equation~\eqref{eq:reality_condition} therefore constitutes a practical guideline: provided the inter-layer couplings moderately exceed the on-site gain-loss contrast, spontaneous emission in a uniformly pumped structure is expected to accumulate preferentially in the helical edge channel, whereas bulk excitations stay marginally stable.  Numerical simulations confirm a broad parameter window in which $\operatorname{Im}E_{\text{edge}}>0$ while $\operatorname{Im}E_{\text{bulk}}=0$.

A conceivable implementation starts from two identical InGaAsP ring-resonator arrays that realise opposite-sign non-Hermitian Chern phases.  Standard photonic-integrated-circuit tooling can, in principle, introduce the required parallel ($\mu$) and anti-parallel ($\nu$) couplings.  Once the inequality~\eqref{eq:reality_condition} is satisfied, uniform electrical or optical pumping should selectively amplify the edge pair without further spatial modulation of the pump.  Crucially, the edge localization mechanism does not rely on microscopic symmetries of the underlying lattice, suggesting a degree of tolerance to fabrication disorder comparable to, or possibly exceeding, that observed in perimeter-pumped devices.

In summary, Eq.~\eqref{eq:reality_condition} furnishes a \emph{condition} for turning the feedback-enforced NH-QSH lattice into a \emph{self-organized topological laser}: bulk suppression + edge amplification emerge from uniform drive, providing the same single-mode, high-efficiency benefits observed experimentally—but with greatly simplified pumping and symmetry requirements.

\section{The self-correcting evolution in the fractal geometry}

In the main text, we have shown that the wavepacket initially localized in the bulk of the stacked-QSH system eventually localizes at the edge, demonstrating self-correcting dynamics. This self-correcting behavior is robust against geometric irregularities, such as those present in fractal geometries.

\begin{figure}[H]
  \centering
  \subfigure[]{\includegraphics[width=0.24\textwidth]{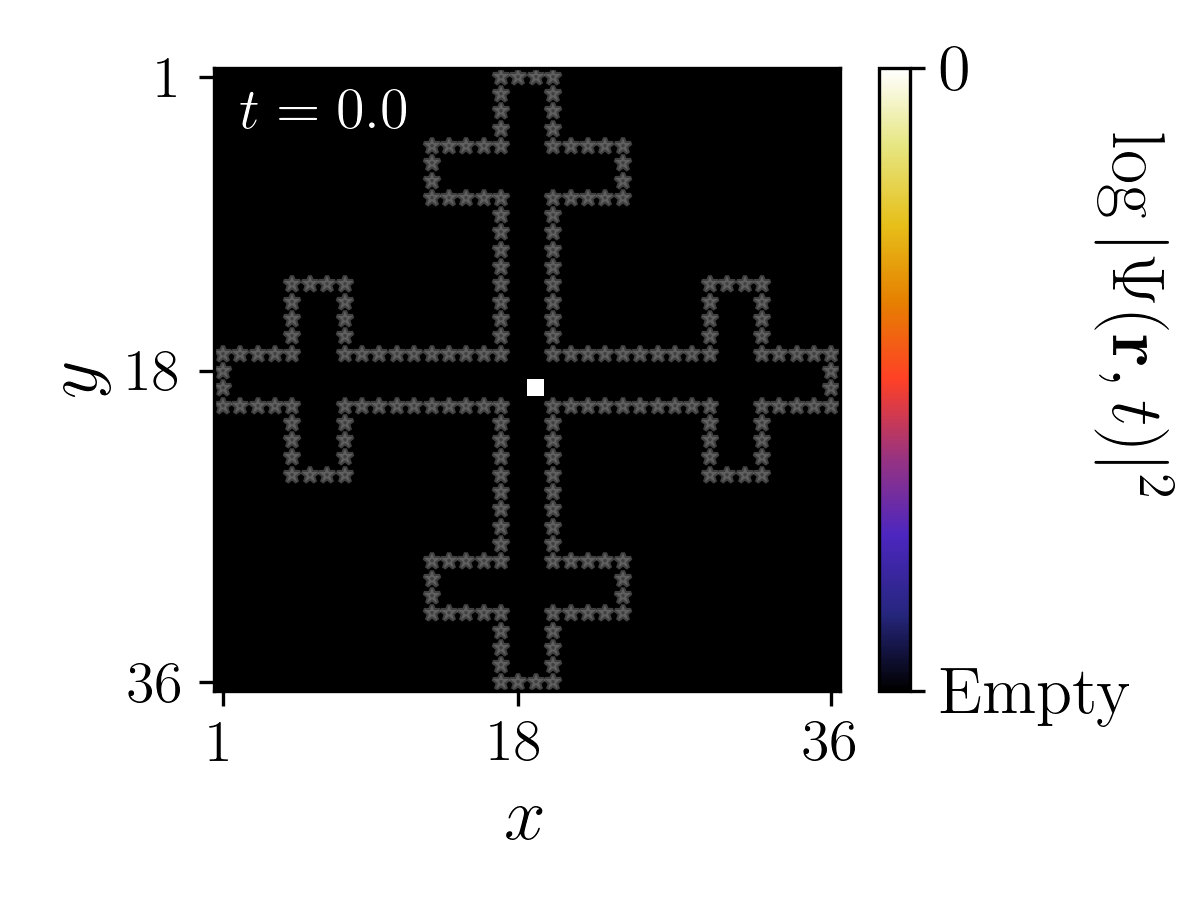}}
  \subfigure[]{\includegraphics[width=0.24\textwidth]{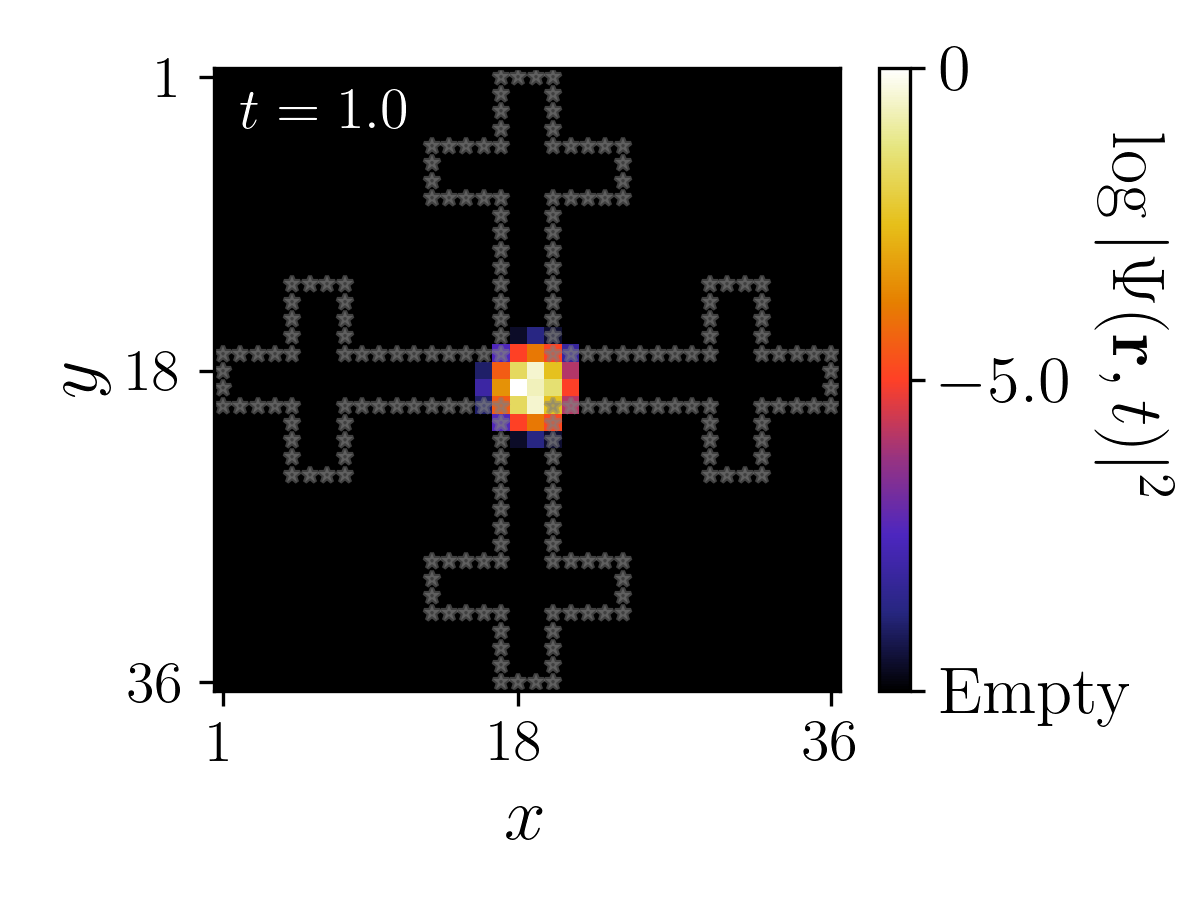}}
  \subfigure[]{\includegraphics[width=0.24\textwidth]{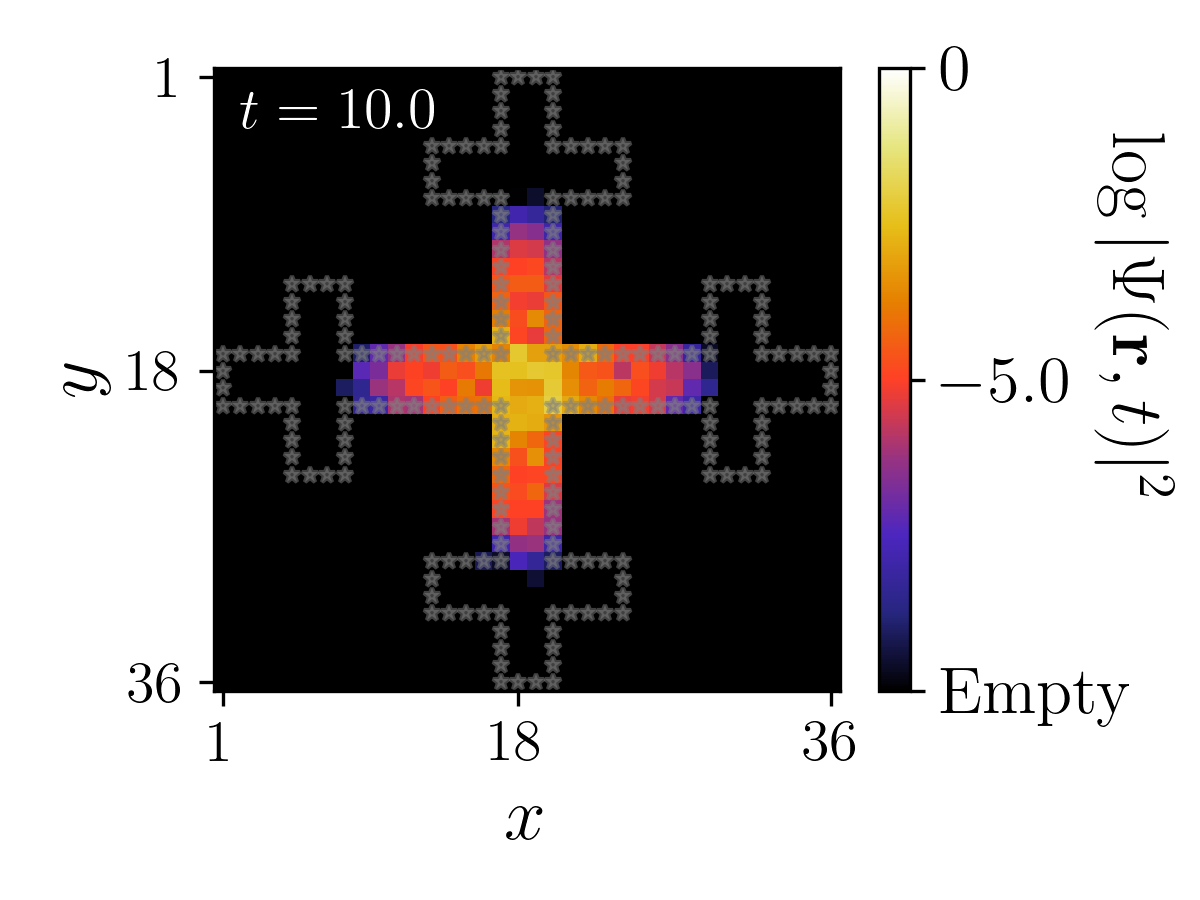}}
  \subfigure[]{\includegraphics[width=0.24\textwidth]{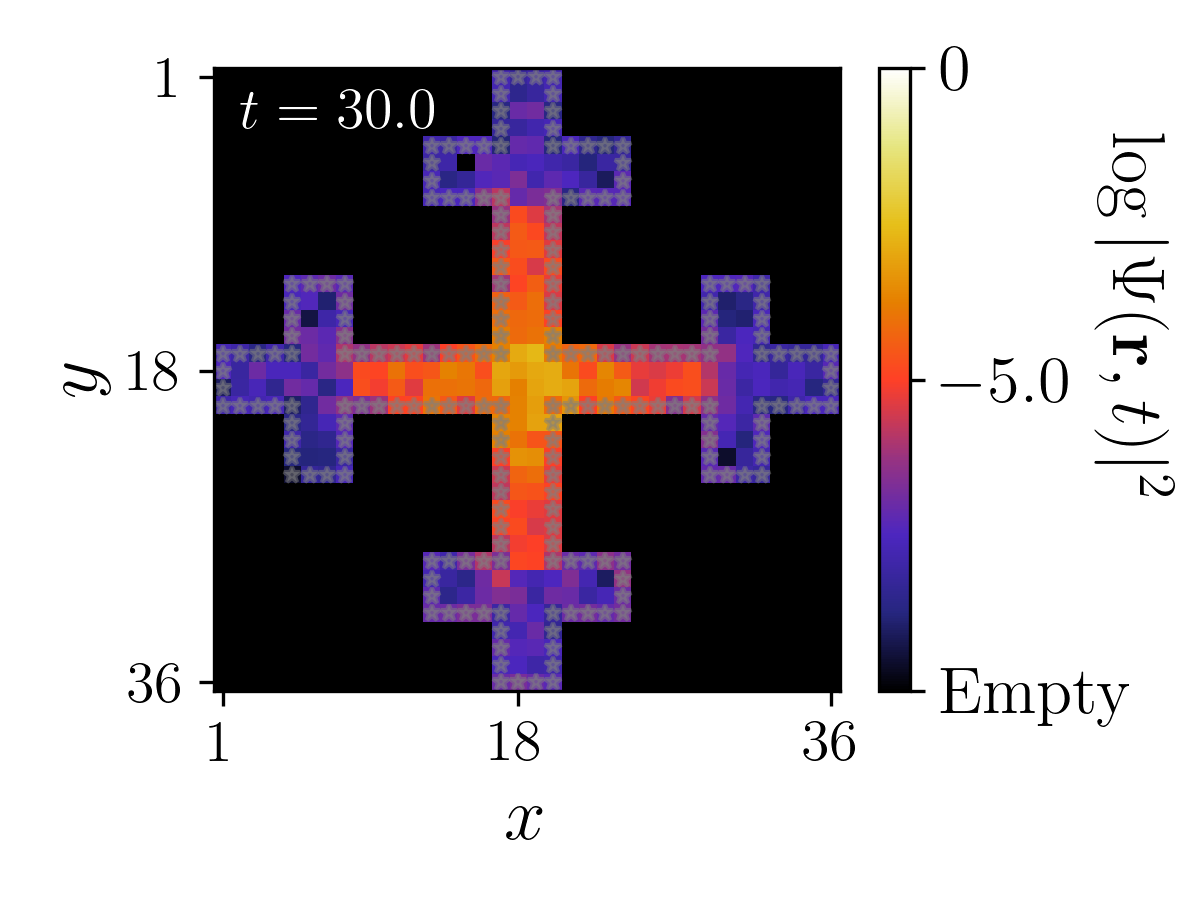}}
  \subfigure[]{\includegraphics[width=0.24\textwidth]{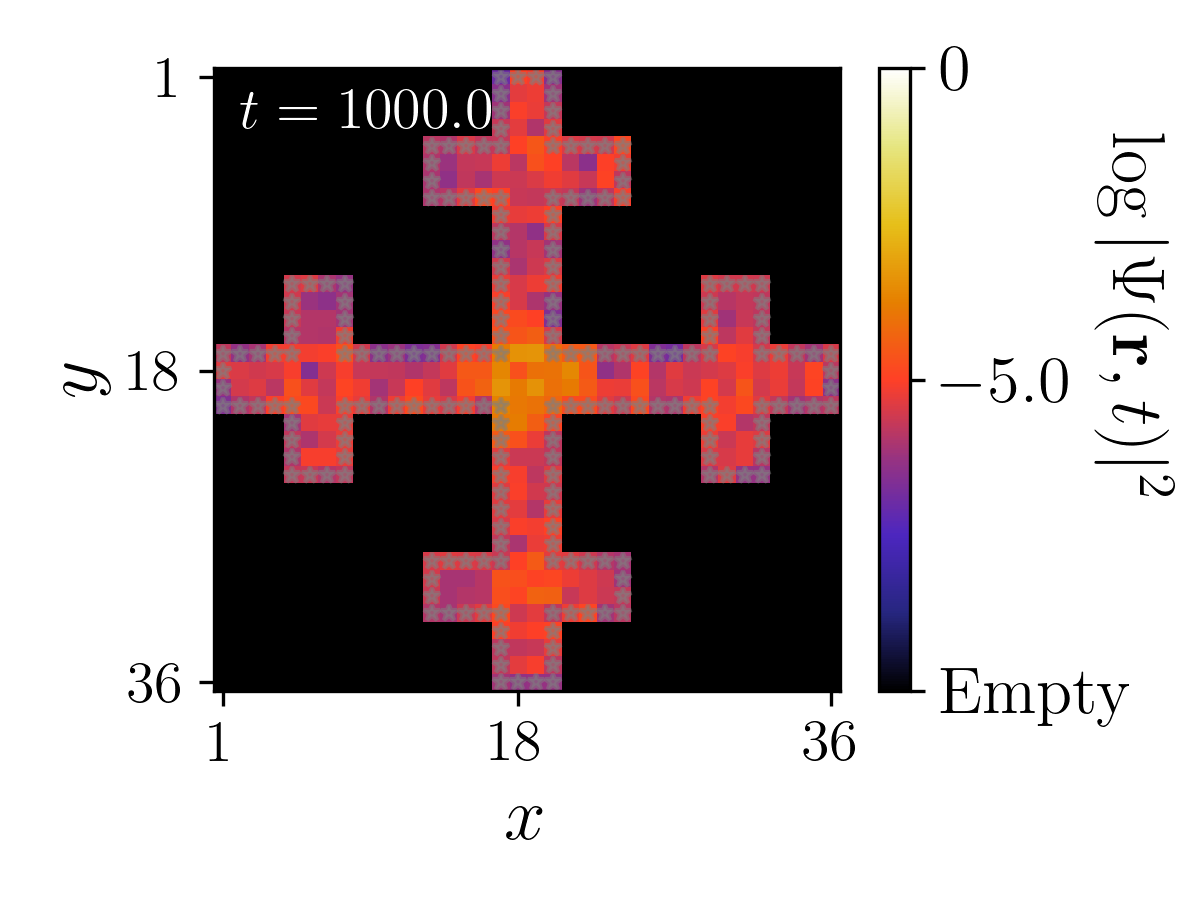}}
  \subfigure[]{\includegraphics[width=0.24\textwidth]{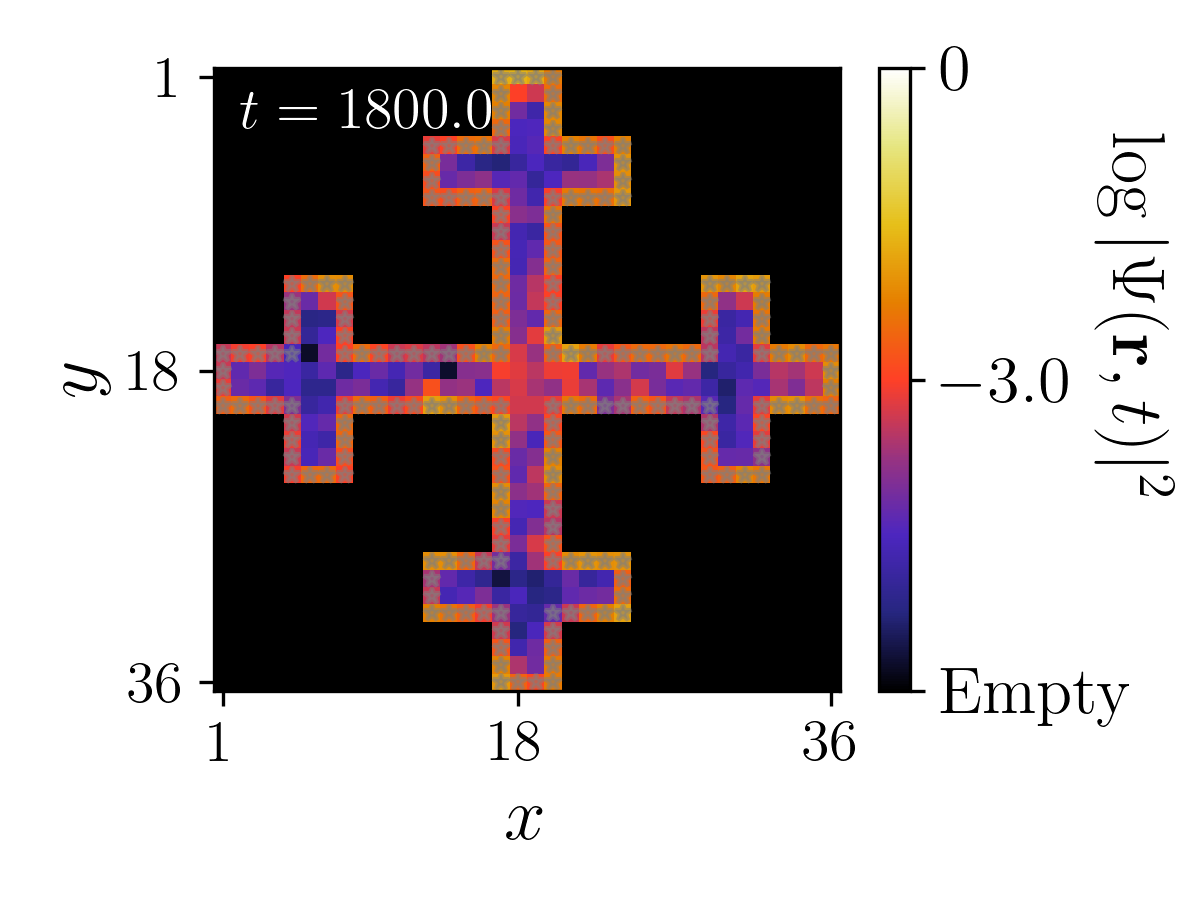}}
  \subfigure[]{\includegraphics[width=0.24\textwidth]{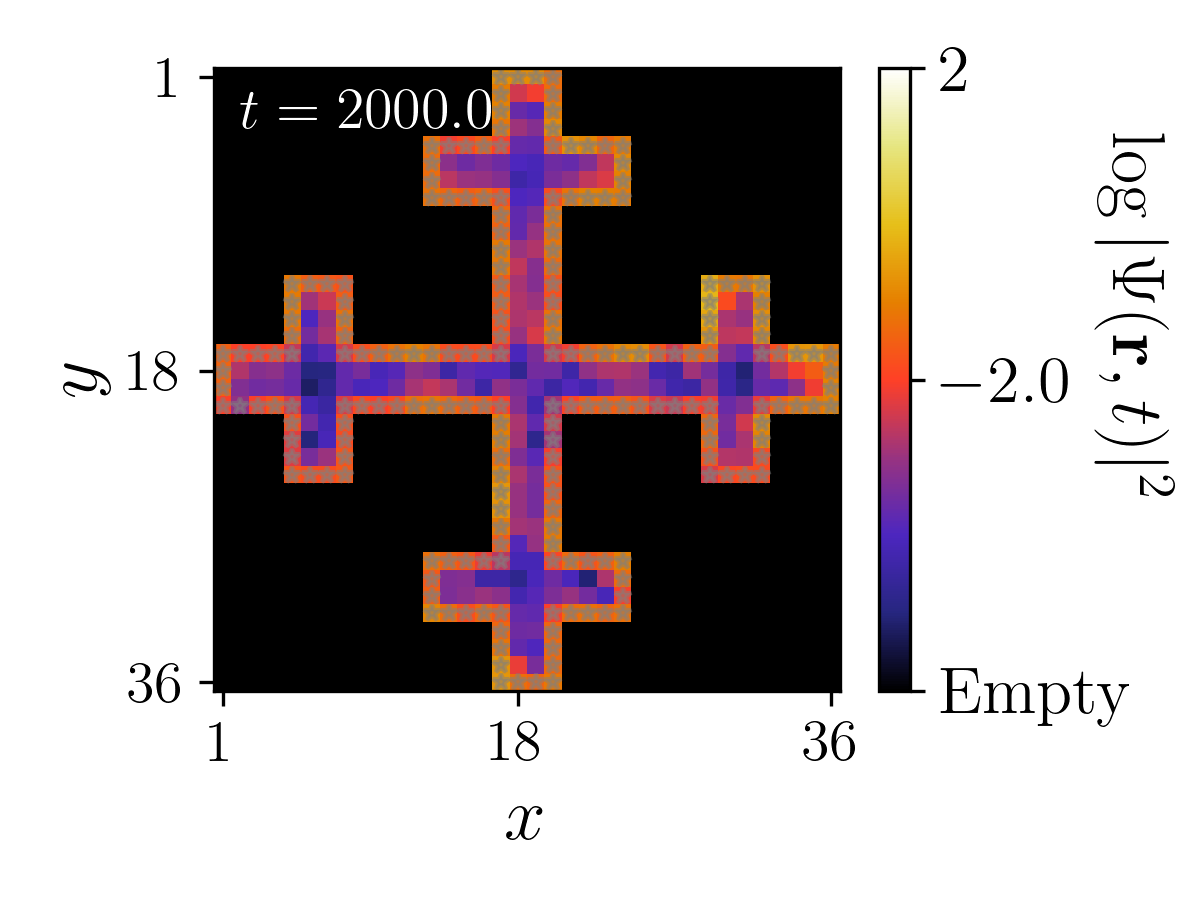}}
  \subfigure[]{\includegraphics[width=0.24\textwidth]{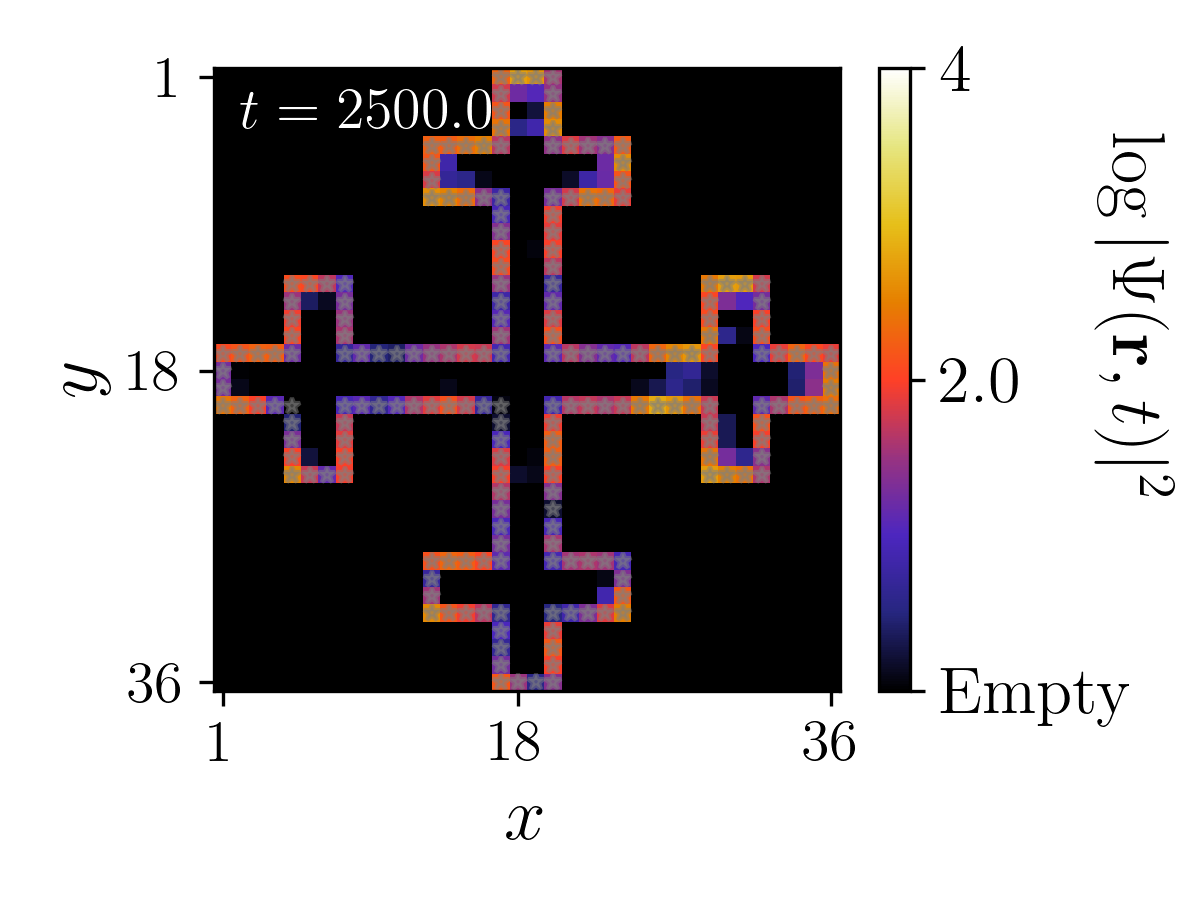}}
  \caption{Density snapshots illustrating the wavepacket propagation in the fractal geometry at various times. Starting from the initial localized bulk state at $t=0$ (a), the wavepacket evolves according to $H_{\text{stacked-QSH}}$, progressively spreading and localizing primarily at fractal boundary regions at later times: (b) $t=1$, (c) $t=10$, (d) $t=30$, (e) $t=1000$, (f) $t=1800$, (g) $t=2000$, (h) $t=2500$. These snapshots complement Fig.~5(a2) in the main text, highlighting robustness against boundary irregularities and clearly illustrating the progressive localization onto fractal edges.}
  \label{fig:fractal_supp}
\end{figure}

The density snapshots shown in Fig.~\ref{fig:fractal_supp} provide a detailed visualization of the wavepacket evolution at selected time instances, directly complementing the edge occupation ratio plot in Fig.~5(a2) from the main text. Initially localized in the bulk, the wavepacket exhibits isotropic spreading at early times, progressively transitioning to predominant localization on the irregular fractal edges at longer times, demonstrating the robustness of the stacked-QSH system against geometric irregularities.

\end{document}